\documentclass[reprint,aps,prl,amsmath,amssymb,amsfonts,dvips]{revtex4-1}
\usepackage[dvips]{graphicx,color}
\usepackage{braket,bm}
\usepackage{dcolumn}
\usepackage{caption}
\begin{document}

\title{
Monte Carlo studies of the spin-chirality decoupling in the three-dimensional Heisenberg spin glass
}
\author{Takumi Ogawa, Kazuki Uematsu and Hikaru Kawamura}
\affiliation{Department of Earth and Space Science, Faculty of Science,
Osaka University, Toyonaka 560-0043,
Japan}
\begin{abstract}
 An extensive equilibrium Monte Carlo simulation is performed on the 3D isotropic Heisenberg SG model with the random nearest-neighbor Gaussian coupling, with particular interest in its chiral-glass (CG) and spin-glass (SG) orderings. For this model, the possibility of the spin-chirality decoupling, {\it i.e.\/}, the CG order setting in at a higher temperature than that of the SG order was suggested earlier, but still remains controversial. We simulate the model up to the maximum size (linear dimension) $L=48$ under both periodic and open boundary conditions (BC). In locating the CG and SG transition temperatures $T_{{\rm CG}}$ and $T_{{\rm SG}}$ by the $L\rightarrow \infty$ extrapolation, a variety of independent physical quantities under the both BC are computed and utilized to get larger number of degrees of freedom (NDF). Thanks to the large NDF up to NDF=43, we succeed in obtaining stable and accurate estimates of the CG and SG transition temperatures, $T_{{\rm CG}}=0.142\pm 0.001$ and $T_{{\rm SG}}=0.131^{+0.001}_{-0.006}$. No sign of the size crossover is observed. For larger $L$, the CG correlation length progressively outgrows the SG correlation length at low temperatures. These results provide strong numerical support for the spin-chirality decoupling. The critical exponents associated with the CG and SG transitions are evaluated by use of the finite-size scaling with the scaling correction. For the CG transition, we get the CG exponents, $\nu_{{\rm CG}}=1.36\pm 0.10$ and $\eta_{{\rm CG}}=0.49\pm 0.10$, consistently with the corresponding experimental exponents of canonical SG. Implications to the chirality scenario of experimental SG ordering is discussed.
\end{abstract}

\maketitle
\section{I. Introduction}
 Spin glasses (SG) have been extensively studied as a prototype of `complex' systems for years, with applications to wide areas like information science, neural networks and deep learning, {\it etc\/}. Originally, the term `spin glass' was coined in the field of magnetism, and has intensively been studied there.  SG as magnets are the type of random magnets possessing both ferromagnetic and antiferromagnetic couplings, and are characterized by frustration and randomness. Due to these effects, SG exhibit nontrivial `glassy' behaviors at low temperatures: see Refs.\cite{Mydoshreview,Kawamurabook} for the review of SG as magnets.

 In as early as 1972, Canella and Mydosh observed that certain SG magnets, especially the so-called canonical SG which are dilute transition-metal - noble-metal alloys, exhibit a sharp cusp-like anomaly suggestive of a thermodynamic transition \cite{Mydosh}. Subsequent experimental studies established that the SG transition is indeed an equilibrium transition and the SG state could be a glassy ordered state in thermal equilibrium, at least in principle \cite{Kawamurabook}. Since then, much experimental and theoretical efforts have been devoted to understand the true nature of the SG transition and the SG ordered state. However, some of the fundamental questions still remain open. 

 As in the case of standard magnets, fundamental properties of the magnetic transition and the magnetic ordered state can be classified by several basic characteristics, {\it e.g.\/}, the space dimensionality $d$ and the order-parameter symmetry, or often in magnets the number of spin components $n$. Ordinary bulk SG are of course three-dimensional ($d=3$) system, so that the number of spin components, or the type of magnetic anisotropy, is expected to be important in their ordering. In many of well-studied experimental SG including canonical SG, the magnetic anisotropy is relatively weak, and can be modelled as an isotropic Heisenberg model with $n=3$-component vector spins, the 3D Heisenberg SG. Of course, even in these Heisenberg-like SG magnets, some amount of anisotropy is inevitable in reality, which might play an important role in real SG ordering. Some other SG possess a significant amount of magnetic anisotropy, which could be either easy-axis-type (Ising-like with $n=1$), or easy-plane-type ($XY\/$-like with $n=2$). Hence, in order to understand the properties of real experimental SG, it would be important to fully understand the ordering properties of the 3D isotropic Heisenberg SG as a reference system. 

 Indeed, the pioneering theoretical work on SG was put forward by Edwards and Anderson (EA) in 1975 \cite{EA}, in which they proposed the model, now standard in the community and called the EA model, which is nothing but the isotropic 3D classical Heisenberg model on the regular lattice with the random nearest-neighbor coupling with both ferromagnetic and antiferromagnetic interactions. EA applied a simple mean-field treatment, while, in order to get fuller understanding of the model properties, one needs to go beyond the MF analysis (the MF analysis itself is already highly nontrivial in SG, though). Since then, several types of numerical simulations have been performed to clarify the ordering properties of the EA model.

 Earlier numerical simulation on the 3D Heisenberg EA model reported in common that the model did not exhibit the SG order at any finite temperature, only the zero-temperature ($T=0$) transition, {\it e.g.\/}, Monte Carlo (MC) simulations by Olive, Young and Sherrington \cite{OYS}, by Matsubara, Iyota and Inawashiro \cite{Matsubara}, and by Yoshino and Takayama \cite{Yoshino}, numerical domain-wall renormalization-group (RG) calculations by Banavar and Cieplak \cite{Banavar}, and by McMillan \cite{McMillan}. It was then argued, {\it e.g.\/}, by Bray, Moore and Young that the experimental finite-temperature SG transition of Heisenberg-like SG was driven solely by the weak magnetic anisotropy inherent to real SG magnets, and the associated anisotropy-driven SG transition belonged to the universality class of the 3D Ising SG \cite{BrayMooreYoung}.

 In 1992, one of the present authors (H.K.) suggested that the model might exhibit a finite-temperature transition in its {\it chiral\/} sector even though the standard SG order did not occur at any finite temperature, proposing the possibility of the {\it spin-chirality decoupling\/}, {\it i.e.\/}, the chirality exhibits the glass transition without accompanying the standard SG order, the chiral-glass (CG) transition, at a finite temperature higher than the standard SG transition temperature \cite{Kawamura1992}. Chirality is a multispin variable, difined locally for three neighboring Heisenberg spins by the scalar $\chi = {\bm S}_i\cdot({\bm S}_j\times {\bm S}_k)$, and takes a nonzero value for the noncoplanar spin configurations with its sign representing the handedness of the noncoplanar spin structure, {\it i.e.\/}, either right- or left-handed.

 On the basis of such a spin-chirality decoupling picture, H.K. proposed the {\it chirality scenario\/} of experimental Heisenberg-like SG magnets \cite{Kawamura1992}. Namely, the true order parameter of real experimental Heisenberg-like SG is the {\it chirality\/} rather than the spin itself, and the properties of the CG transition and the CG order, which is `hidden' in the chirality in the hypothetical spin-chirality-decoupled fully isotropic system, is `revealed' in the spin in real Heisenberg-like SG via the weak random magnetic anisotropy. The chirality scenario was further extended since then: see Refs.\cite{Kawamurabook,Kawamurareview} for details. Concerning the spin-chirality decoupling in the fully isotropic 3D Heisenberg SG, which forms a basis of the chirality scenario, some support was subsequently reported from MC \cite{Kawamura1995,Kawamura1998,HukushimaKawamura2000,Matsumoto}. In particular, Hukushima and Kawamura reported that the CG order exhibited a one-step-like peculiar replica-symmetry breaking (RSB) \cite{HukushimaKawamura2000}, quite different from the full-step or hierarchical RSB discussed in connection with the 3D Ising SG \cite{Mydoshreview,Kawamurabook}. 

 Concerning the question of the standard SG order setting in either at $T=0$ or $T>0$, the view of the community has changed since then, mainly because the progress in the computing capability enabled one to look into the low-temperature region in more details. Namely, there now seems to be a consensus that the SG order of the 3D isotropic Heisenberg SG sets in at a finite temperature, {\it i.e.\/}, $T_{{\rm SG}}>0$, in contrast to the earlier belief of $T_{{\rm SG}}=0$ \cite{OYS,Matsubara,Yoshino,Banavar,McMillan}. However, whether there occurs the spin-chirality decoupling or not, {\it i.e.\/}, whether $T_{{\rm CG}}>T_{{\rm SG}}>0$ or $T_{{\rm CG}}=T_{{\rm SG}}>0$, is still at issue.

 On the basis of the MC simulation for relative small sizes of the linear dimension $L\leq 12$, Lee and Young claimed that the model exhibited a simultaneous spin and chiral transition at $T_{{\rm SG}}/J=0.16\pm 0.02$, so no spin-chirality decoupling  ($J$ is the standard deviation of the Gaussian distribution for the coupling) \cite{LeeYoung2003}. By contrast, Hukushima and Kawamura suggested \cite{HukushimaKawamura2005}, by simulating the binary-coupling ($\pm J$) model of $L\leq 20$ (a different coupling model from those treated in Refs.\cite{LeeYoung2003,Campos,CampbellKawamura,LeeYoung2007,VietKawamuraPRL,VietKawamuraPRB,Fernandez,Nakamura}), that the spin and the chirality were decoupled, {\it i.e.\/}, while $T_{{\rm SG}}$ was either zero or nonzero but less than $T_{{\rm CG}}$, {\it i.e.\/}, $T_{{\rm SG}} < T_{{\rm CG}}$. They also argued that such a decoupling was visible clearly only on the length scale exceeding a certain crossover length scale of $\sim 20$, simply because the chirality is locally defined as a composite operator of the spin variables \cite{HukushimaKawamura2005}. Campos {\it et al\/} claimed, by extending the maximum system size up to $L=32$ in their MC simulation on the Gaussian-coupling model, that the model exhibited a simultaneous spin and chiral transition of the Kosterlitz-Thouless (KT) type, the system lying close to the lower critical dimension  (no explicit report of the $T_{{\rm SG}}(=T_{{\rm CG}}$)-value was given) \cite{Campos}. Criticism to such an interpretation was subsequently given in Ref.\cite{CampbellKawamura}, however. Lee and Young also studied the lattice up to $L=32$ for the same model, and observed a marginal behavior for larger sizes, claiming a simultaneous spin and chiral transition (no explicit report of the $T_{{\rm SG}}(=T_{{\rm CG}}$)-value given, though) \cite{LeeYoung2007}. By contrast, Viet and Kawamura claimed on the basis of the MC simulation on the same model with $L\leq 32$ that the model exhibited the spin-chirality decoupling \cite{VietKawamuraPRL,VietKawamuraPRB}, the estimated transition temperatures being $T_{{\rm CG}}=0.143\pm 0.003$ and $T_{{\rm SG}}=0.125^{+0.001}_{-0.005}$ \cite{VietKawamuraPRB}. These authors also estimated the critical exponents associated with the CG transition, {\it i.e.\/}, the CG correlation-length exponent $\nu_{{\rm CG}}= 1.4\pm 0.2$ and the CG critical-point-decay exponent $\eta_{{\rm CG}}=0.6 \pm 0.2$, which turned out to be rather close to the experimentally observed exponents of canonical SG \cite{Kawamurabook}. This agreement favors the chirality scenario, since in the chirality scenario, the SG exponents of real Heisenberg-like SG should be those of the CG exponents of the fully isotropic Heisenberg SG. Subsequent MC simulation by Fernandez {\it et al\/} on the same model for $L\leq 48$, however, suggested that the spin and the chirality ordered at the same temperature $T_{{\rm SG}}=T_{{\rm CG}}=0.120^{+0.010}_{-0.100}$ \cite{Fernandez}. Furthermore, the recent nonequilibrium MC study by Nakamura on the same model reported a simultaneous spin and chiral transition at $T_{{\rm SG}}=T_{{\rm CG}}=0.140\pm 0.002$ \cite{Nakamura}. 

 Those MC simulations on the 3D isotropic Heisenberg SG model were performed under periodic boundary conditions (BC). In this connection, an interesting work was made by Shirakura and Matsubara who examined the effect of finite sizes by imposing a different type of BC, {\it i.e.\/}, open BC, on the same model, though they did not reach any conclusion concerning the occurrence of the spin-chirality decoupling \cite{Shirakura}. 

 Evidently, the present numerical situation on the 3D Heisenberg SG model, especially the occurrence of the spin-chirality decoupling, remains entangled and controversial. For example, if one cites the works claiming the absence of the spin-chirality decoupling, some did not give an explicit estimate of $T_{{\rm SG}}(=T_{{\rm CG}}$)-value, while even when reported, the quoted simultaneous spin and chiral transition temperatures are distributed as $T_{{\rm SG}}(=T_{{\rm CG}})=0.16\pm 0.02$ \cite{LeeYoung2003}, $T_{{\rm SG}}=T_{{\rm CG}}=0.120^{+0.010}_{-0.100}$ \cite{Fernandez} and $T_{{\rm SG}}=T_{{\rm CG}}=0.140\pm 0.002$ \cite{Nakamura}. In fact, the higher one came rather close to, or even higher than the chiral $T_{{\rm CG}}$ estimate of Ref.\cite{VietKawamuraPRB}, $T_{{\rm CG}}=0.143\pm 0.003$, claiming the spin chirality decoupling, and the lower one came close to that of the spin $T_{{\rm SG}}$ estimate of Ref.\cite{VietKawamuraPRB}, $T_{{\rm SG}}=0.125^{+0.006}_{-0.012}$. Obviously, purely from the numerical viewpoint, the situation needs to be further clarified.

 Furthermore, the chirality scenario based on the spin-chirality decoupling picture of the 3D isotropic Heisenberg SG has been rather successful in explaining certain issues of experimental SG, including the issue of the criticality and the magnetic phase diagram \cite{ImagawaKawamura2001,ImagawaKawamura2004,PetitCampbell,Kawamurareview,Kawamurabook}. The scenario also has got a rather direct experimental support from the Hall measurements on canonical SG \cite{Taniguchi2004,Campbell2004,Campbell2006,Taniguchi2007,Yamanaka2007} probing the chiral response of the SG \cite{TataraKawamura,KawamuraHall}. In view of such a promising status of the chirality scenario, it would be important to further clarify the issue of the spin-chirality decoupling in the fully isotropic 3D Heisenberg SG model.

 Under such circumstances, we undertake in the present paper a new set of MC simulation on the 3D isotropic Heisenberg SG model with the random nearest-neighbor Gaussian coupling, exactly the same model as studied previously by many authors. Our maximum size is $L=48$, the same as simulated in Ref.\cite{Fernandez}. However, we simulate both periodic and open BC in parallel, and utilize the both data simultaneously in locating the transition temperatures $T_{{\rm CG}}$ and $T_{{\rm SG}}$. In addition, we compute and utilize a variety of independent physical quantities, not only the crossing temperatures of the correlation-length ratio under periodic BC utilized in Refs.\cite{Campos,LeeYoung2003,LeeYoung2007,Fernandez}, but also those under open BC, and the crossing temperatures and the dip temperatures of the Binder ratio as well. Our strategy is to utilize as many independent information (data points) as possible to get larger number of degrees of freedom (NDF) in the necessary size extrapolation to the $L\rightarrow \infty $ limit in order to reduce and control the error bar. Indeed, we can get the NDF as large as 43. Making use of the obtained large NDF, we carefully examine the stability of our estimates of $T_{{\rm CG}}$ and $T_{{\rm SG}}$.

 Finally, we succeed in obtaining rather stable and accurate estimates of the CG and SG transition temperatures as $T_{{\rm CG}}=0.142\pm 0.001$ and $T_{{\rm SG}}=0.131^{+0.001}_{-0.006}$. The results provide strong numerical support for the spin-chirality decoupling. We also determine the critical exponents associated with the CG and SG transitions. For the CG transition, we get the CG exponents, $\nu_{{\rm CG}}=1.36\pm 0.10$ and $\eta_{{\rm CG}}=0.49\pm 0.10$, which are consistent with the earlier reports and are also consistent with the corresponding experimental values on canonical SG, $\nu\simeq 1.3-1.4$ and $\eta\simeq 0.4$ \cite{Kawamurabook}. This agreement gives support to the chirality scenario of the experimental SG ordering. The one-step-like feature of the CG ordering reported earlier is also confirmed for larger sizes than before.

 The rest of the present paper is organized as follows. In \S II, we introduce our model and the method employed. Section III is the main part of the present paper where we present our Monte Carlo results. We first define various physical quantities computed in the present paper in \S IIIA. In \S IIIB, we present our MC data of the CG and SG correlation-length ratios and the CG and SG Binder ratios for both cases of periodic and open BC. Making full use of these data, we estimate the CG and SG transition temperatures in \S IIIC and D, respectively. The interrelation between the CG correlation length $\xi_{{\rm CG}}$  and the SG correlation length $\xi_{{\rm SG}}$  is examined in \S IIIE, with particular interest in the relative magnitude of $\xi_{{\rm CG}}$ and $\xi_{{\rm SG}}$. The MC data of the chiral overlap distribution is presented in \S IIIF. In \S IV, we analyze the critical properties of the CG and SG transitions in \S IVA and IVB, respectively, by use of the finite-size scaling taking account of the scaling correction. Various CG and SG critical exponents are estimated. Finally, section V is devoted to summary and discussion.

\section{II. The model and the method}
We study an isotropic classical Heisenberg model on a 3D simple-cubic lattice whose Hamiltonian is given by
\begin{equation}
{\cal H}=-\sum_{<ij>}J_{ij}{\bm S}_i\cdot {\bm S}_j\ \ ,
\label{eqn:hamil}
\end{equation}
where ${\bm S}_i=(S_i^x,S_i^y,S_i^z)$ ($|{\bm S}_i|=1$) is a three-component unit vector at the $i$-th site, and the sum over $<ij>$ is taken over all nearest-neighbor pairs on the lattice. The couplings $J_{ij}$ are the random Gaussian variables with the mean zero and the variance unity. We apply the two types of BC, {\it i.e.\/}, i) periodic BC in all three directions, and (ii) open BC  in all three directions.  The lattice contains $N=L^{3}$ sites, where the lattice linear dimension is $L=6, 8, 12, 16, 20, 24, 32, 40, 48$ for periodic BC, and $L=6, 8, 12, 16, 20, 24, 32, 40$ for open BC.
\begin{center}
 \begin{table}
\captionsetup{labelformat=empty,labelsep=none}
\caption{[Periodic boundary conditions]}
  \begin{tabular*} {0.5\textwidth} {@{\extracolsep{\fill}} c c c c c c}
   \hline
   \hline
   $L$ & $N_{s}$ & $N_{T}$ & $N_{{\rm MC}}$ & $T_{{\rm max}}$ & $T_{{\rm min}}$ \\ 
   \hline
   6 & 2000 & 32 & $1\times10^5$ & 0.333 & 0.111 \\
   8 & 2000 & 32 & $1\times10^5$ & 0.333 & 0.111 \\
   12 & 2000 & 32 & $1\times10^5$ & 0.333 & 0.111 \\
   16 & 2000 & 32 & $1\times10^5$ & 0.222 & 0.121 \\
   20 & 1500 & 44 & $1\times10^5$ & 0.209 & 0.121 \\
   24 & 1500 & 44 & $1.5\times10^5$ & 0.209 & 0.121 \\
   32 & 344 & 56 & $8\times10^5$ & 0.196 & 0.12583 \\
   40 & 320 & 72 & $16\times10^5$ & 0.196 & 0.12583 \\
   48 & 176 & 96 & $24\times10^5$ & 0.190 & 0.12794 \\
   \hline
   \hline
\smallskip
  \end{tabular*}
%\captionsetup{labelformat=empty,labelsep=none}
\caption{[Open boundary conditions]}
  \begin{tabular*} {0.5\textwidth} {@{\extracolsep{\fill}} c c c c c c}
   \hline
   \hline
   $L$ & $N_{s}$ & $N_{T}$ & $N_{{\rm MC}}$ & $T_{{\rm max}}$ & $T_{{\rm min}}$ \\ 
   \hline
   6 & 2000 & 32 & $1\times10^5$ & 0.250 & 0.08 \\
   8 & 2000 & 32 & $1\times10^5$ & 0.250 & 0.08 \\
   12 & 2000 & 32 & $1\times10^5$ & 0.250 & 0.08 \\
   16 & 2000 & 32 & $1\times10^5$ & 0.250 & 0.099 \\
   20 & 288 & 48 & $1\times10^5$ & 0.200 & 0.111 \\
   24 & 1000 & 48 & $1\times10^5$ & 0.200 & 0.111 \\
   32 & 401 & 48 & $4\times10^5$ & 0.185 & 0.111 \\
   40 & 192 & 72 & $16\times10^5$ & 0.185 & 0.11468 \\
   \hline
   \hline
\smallskip
  \end{tabular*}
\caption{TABLE 1.  Various parameters of the present Monte Carlo simulation.  $L$ is the system size (the lattice linear dimension), $N_{s}$ is the number of samples, $N_{{\rm MC}}$ is the total number of Monte Carlo steps per spin (our unit Monte Carlo step consists of 1 heat-bath sweep and $L$ over-relaxation sweeps),  $T_{{\rm max}}$ and $T_{{\rm min}}$ are the highest and lowest temperatures used in the temperature-exchange run, and $N_{T}$ is the total number of temperature points. Measurements of physical quantities are made over the latter half of the total $N_{{\rm MC}}$ Monte Carlo steps, while the former half is discarded for thermalization.}
\label{table1}
\end{table}
\end{center}

 Thermodynamic properties of the model are computed by means of MC simulation based on the standard heat-bath method and the over-relaxation method, which are combined with the temperature-exchange technique \cite {HukushimaNemoto}. One MC step per spin (MCS) consists of one heat-bath sweep followed by $L$ successive over-relaxation sweeps. After every MC step, we perform the temperature-exchange trial, which is made between the two spin configurations at a pair of neighboring temperatures. In the temperature range between $T_{{\rm min}}$ and $T_{{\rm max}}$, $N_{T}$ distinct temperatures are distributed so that the acceptance rate of the exchange trial takes a moderate value, say, greater than $\sim 0.1$. The maximum temperature $T_{{\rm max}}$ is chosen to be high enough so that the autocorrelation time by the single-spin-flip dynamics is short enough. 

 In Table I, we show some of the details of our simulation conditions, including the system size (linear dimension) $L$, the number of independent samples (bond realizations) $N_{s}$ , the number of temperature points used in the temperature-exchange process $N_{T}$, the minimum and maximum temperatures $T_{{\rm min}}$ and $T_{{\rm max}}$, and the total MCS performed per replica. The measurement is made over the latter half of $N_{{\rm MC}}$ MCS, while the former half is discarded for thermalization. The initial spin configuration is taken to be random.

Error bars are estimated via sample-to-sample fluctuations for linear quantities like the order parameter, and by the bootstrap method for non-linear quantities like the Binder ratio and the correlation-length ratio.

 To ensure full thermalization of the system is crucially important. In particular, special attention needs to be paid to thermalize the chirality-related quantities at low temperatures, as the chirality is an Ising-like discrete quantity ({\it i.e.\/}, right or left) with a finite energy barrier to be overcome to flip it. This point is crucially important in the present task since the possible poor or insufficient thermalization of the chirality-related quantities, if it happened, would apparently weaken the ordering tendency of the chirality, and obscure the spin-chirality decoupling. In the present work, we follow Ref.\cite{VietKawamuraPRB} to check the thermalization by carefully observing the conditions 1)-5) of Ref.\cite{VietKawamuraPRB}. Namely, 1) all of the `replicas' move back and forth many times along the temperature axis during the temperature-exchange process (typically more than 10 times) between the maximum and minimum temperature points, with sufficintly fast relaxation achieved even without the temperature exchange process at the higherst temperature; 2) the thermodynamic relation among the energy, the `link overlap' and the `spin overlap' expected to hold for the model with the Gaussian bond distribution in equilibrium is satisfied; 3) measured physical quantities converge to stable values as a function of the MC time; 4) the expected symmetry of the overlap distribution function holds for each individual sample; and 5) the equality between the specific heat computed via the energy fluctuation and the one computed via the temperature difference of the energy holds.

\section{III. Monte Carlo results}

In this section, we present our MC results, with a focus on the issue of the possible spin-chirality decoupling, {\it i.e.\/}, whether $T_{{\rm CG}}$ and $T_{{\rm SG}}$ are different or common. For this purpose, we concentrate on the two kinds of dimensionless quantities in the following, {\it i.e.\/}, the correlation-length ratio $\xi /L$ and the Binder ratio $g$ both for the spin and the chirality.

\subsection{A. Physical quantities}

 In this subsection, we first give the definitions of the physical quantities we compute. Let us begin with the chirality-related quantities. The local chirality  $\chi_{i\mu}$ at the $i$-th site and in the $\mu$-th ($\mu=x,y,z$) direction is defined for three neighboring Heisenberg spins on a line by
\begin{equation}
\chi_{i\mu}=
{\bm S}_{i+{\hat{e}}_{\mu}}\cdot
({\bm S}_i\times {\bm S}_{i-{\hat{e}}_{\mu}}),
\end{equation}
where ${\hat{e}}_{\mu}$ denotes a unit vector along the $\mu$-th axis. In the present definition of the chirality, we consider the three spins (spin-triad) on a line. Let the total number of independent spin-triads $N_t$. $N_t=3N$ in case of periodic BC, while $N_t=3L^2(L-2)=3N-6L^2$ in case of open BC.

 By considering the two `replicas', {\it i.e.\/}, two independent systems 1 and 2 with the same bond realization \{$J_{ij}$\}, the chiral overlap $q_\chi$ may be defined by
\begin{equation}
q_\chi = \frac{1}{N_t}\sum_{{\rm triad}}
\chi_{i\mu}^{(1)}\chi_{i\mu}^{(2)},
\end{equation}
where the summation is taken over all independent spin-triads for which the local chirality is defined. In actual simulations, we simulate these two replicas 1 and 2 in parallel with using different spin initial conditions and different random-number sequences.

 The CG susceptibility $\chi_{{\rm CG}}$ might be defined via the second moment of $q_\chi$ by
\begin{equation}
\chi_{{\rm CG}} = N_t \left[ \langle q_\chi^2\rangle \right],
\end{equation}
where $\langle \cdots \rangle$ represents the thermal average and $[\cdots ]$ the average over the bond disorder $\{ J_{ij}\}$. The CG Binder ratio is defined by
\begin{equation}
g_{{\rm CG}}=
\frac{1}{2}
\left(3-\frac{[\langle q_{\chi}^4\rangle]}
{[\langle q_{\chi}^2\rangle]^2}\right).
\end{equation}

The chiral overlap may be extended to the $k$-dependent quantity as
\begin{equation}
q_\chi({\bm k}) =
\frac{1}{N_t}\sum_{{\rm triad}}
\chi_{i\mu}^{(1)}\chi_{i\mu}^{(2)}e^{i{\bm k}\cdot {\bm r}_i}.
\label{chioverlap}
\end{equation}
Then, the finite-size CG correlation length $\xi_{{\rm CG}}$ is defined by
\begin{equation}
\xi_{{\rm CG}} = 
\frac{1}{2\sin(k_\mathrm{m}/2)}
\sqrt{ \frac{ [\langle q_\chi({\bm 0})^2 \rangle] }
{[\langle q_\chi({\bm k}_\mathrm{m})^2 \rangle] } -1 },
\label{CGxi}
\end{equation}
where ${\bm k}_{\rm m}=(2\pi/L,0,0)$ and $k_{\textrm{m}}=|{\bm k}_{\textrm{m}}|$. We restrict the $\mu$-direction in Eqs.(\ref{chioverlap}) and (\ref{CGxi}) to be parallel with ${\bm k}$.

 For the Heisenberg spin, the overlap becomes a tensor varialbe in the spin space, and an appropriate $k$-dependent spin overlap may be defined by
\begin{equation}
q_{\alpha\beta}({\bm k}) = 
\frac{1}{N}\sum_{i=1}^N S_{i\alpha}^{(1)}S_{i\beta}^{(2)}e^{i{\bm k}\cdot {\bm r}_i},
\ \ \ (\alpha,\beta=x,y,z).
\end{equation}
The associated SG susceptibility is defined by
\begin{equation}
\chi_{{\rm SG}} = N[\langle q_{\rm s}({\bm 0})^2\rangle]\ ,
\ \ \  
q_{\rm s}({\bm k})^2 = \sum_{\alpha,\beta=x,y,z} \left| q_{\alpha\beta}({\bm k}) \right| ^2.
\end{equation}
The finite-size SG correlation length is defined by
\begin{equation}
\xi_{{\rm SG}} = 
\frac{1}{2\sin(k_\mathrm{m}/2)}
\sqrt{ \frac{ [\langle q_s({\bm 0})^2 \rangle] }
{[\langle q_s({\bm k}_\mathrm{m})^2 \rangle] } -1 } .
\end{equation}
 The CG and SG correlation-length ratios are then defined by $\xi_{{\rm CG}} /L$ and $\xi_{{\rm SG}} /L$. 

 The SG Binder ratio is defined by
\begin{equation}
g_{{\rm SG}} = \frac{1}{2}
\left(11 - 9\frac{[\langle q_{\rm s}({\bm 0})^4\rangle]}
{[\langle q_{\rm s}({\bm 0})^2\rangle]^2}\right).
\label{eqn:gs_def}
\end{equation}
 The SG and CG Binder ratios are normalized so that, in the thermodynamic limit, they vanish in the high-temperature phase and gives unity in the non-degenerate ordered state. In the present Gaussian coupling model, the ground state is expected to be non-degenerate so that both $g_{{\rm CG}}$ and $g_{{\rm SG}}$ should take a value unity at $T=0$.

\subsection{B. The Monte Carlo data}

 Now, we present our MC data. First, we show in Fig.1 the temperature and size dependence of the correlation-length ratio (a) for the chirality, and (b) for the spin, in the case of periodic BC. The corresponding figures for open BC are given in Figs.2(a) and (b), respectively. In both Figs.1 and 2, main figures exhibit the data in the transition region, while the wider temperature range is covered in the insets.

 General characteristic of the correlation-length ratio $\xi/L$ is that the data for different $L$ cross with each other with its crossing temperature converging to the bulk $T_c$ in the  $L\rightarrow \infty$ limit. As can clearly be seen from the figures, there occurs a crossing between different size data both in Figs.1 and 2, indicative of a phase transition in the chirality and spin sectors. To identify the bulk CG and SG transition temperature $T_{{\rm CG}}$ and $T_{{\rm SG}}$, one needs to perform an appropriate extrapolation of finite-$L$ crossing temperatures $T_{{\rm cross}}(L)$ to the thermodynamic limit $L\rightarrow \infty$.

\begin{figure}[ht]
\begin{center}
\includegraphics[width=\hsize]{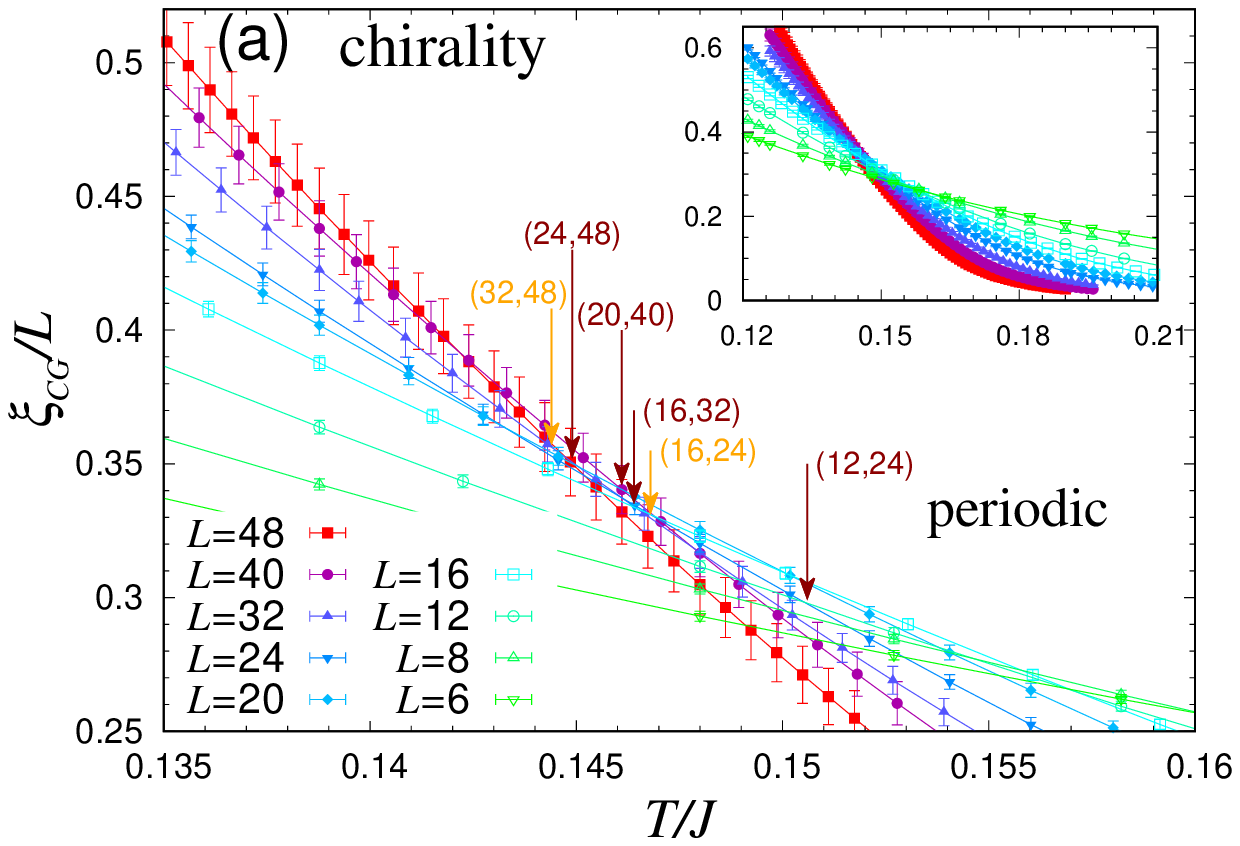}
\includegraphics[width=\hsize]{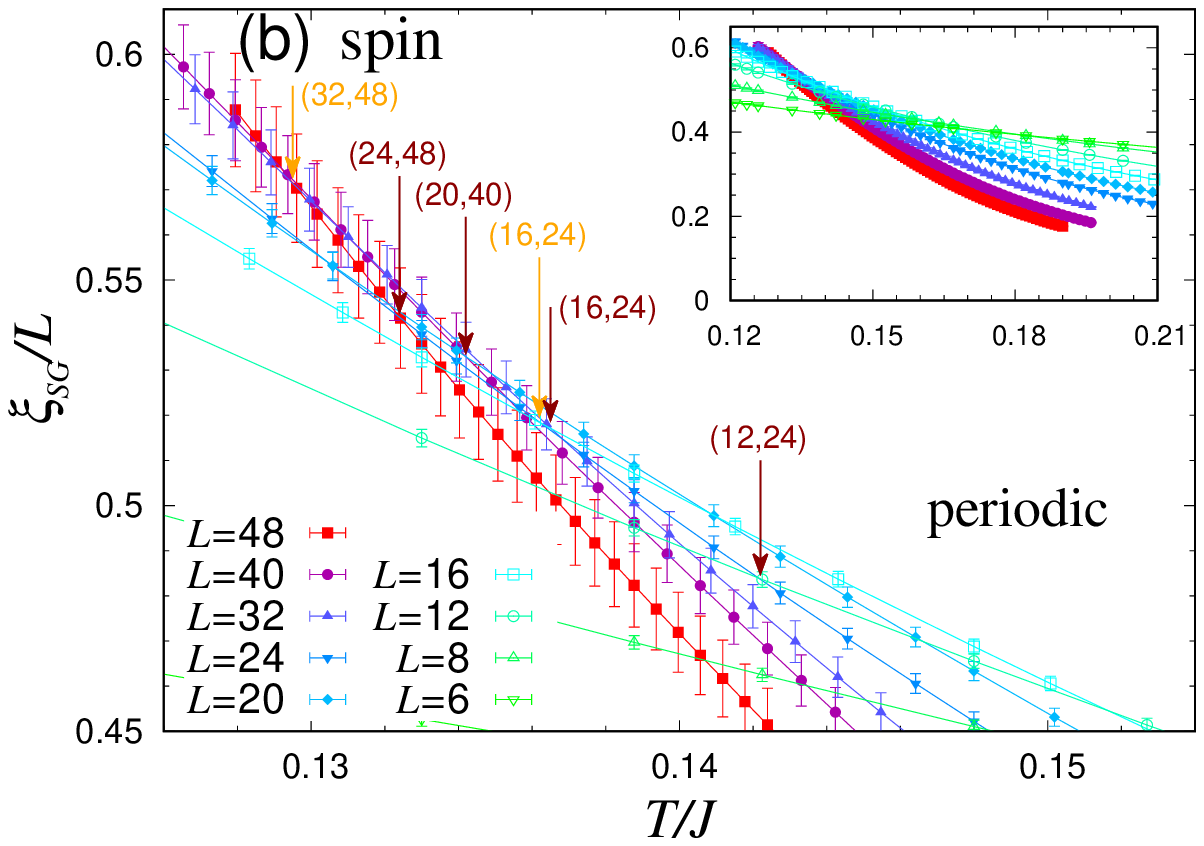}
\end{center}
\caption{
(Color online) The temperature and size dependence of the correlation-length ratio under periodic BC, (a) for the chirality, and (b) for the spin. Main panel is a magnified view of the transition region, while the inset represents a wider temperature region.
}
\end{figure}
\begin{figure}[ht]
\begin{center}
\includegraphics[width=\hsize]{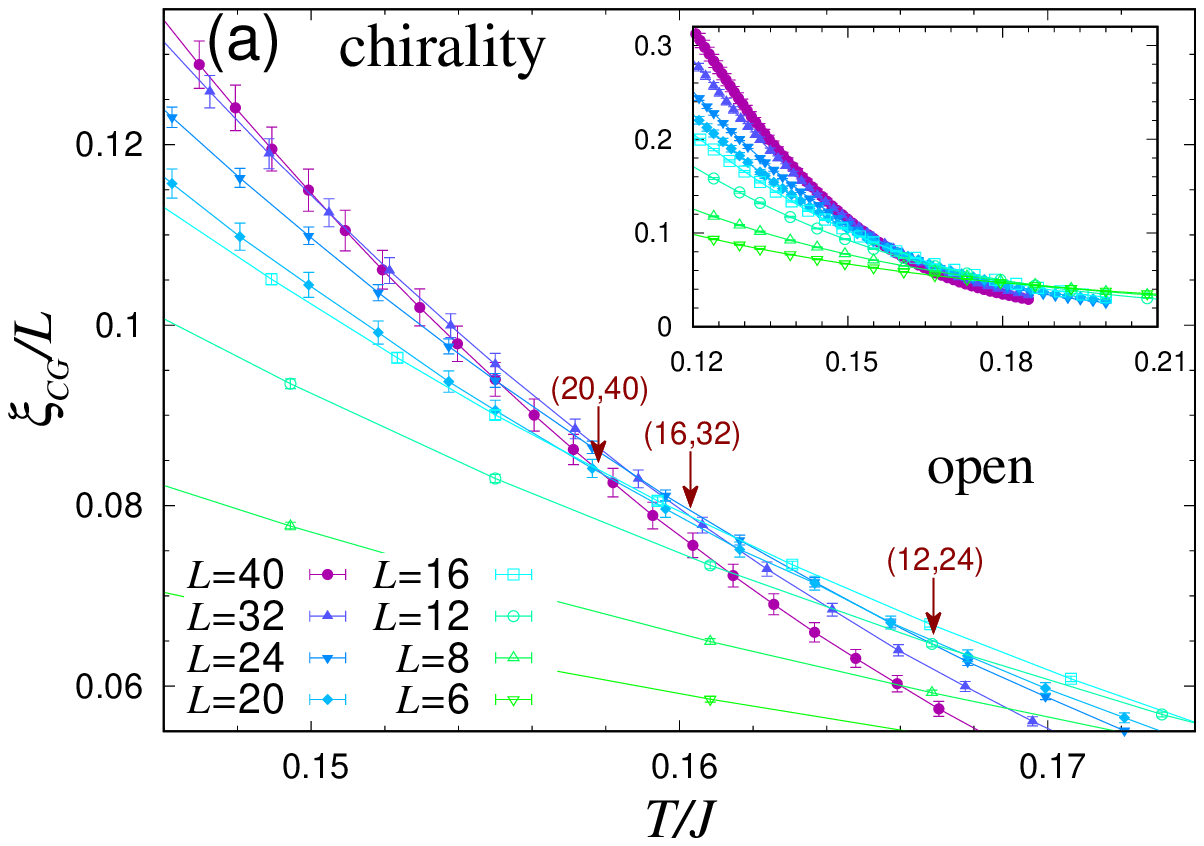}
\includegraphics[width=\hsize]{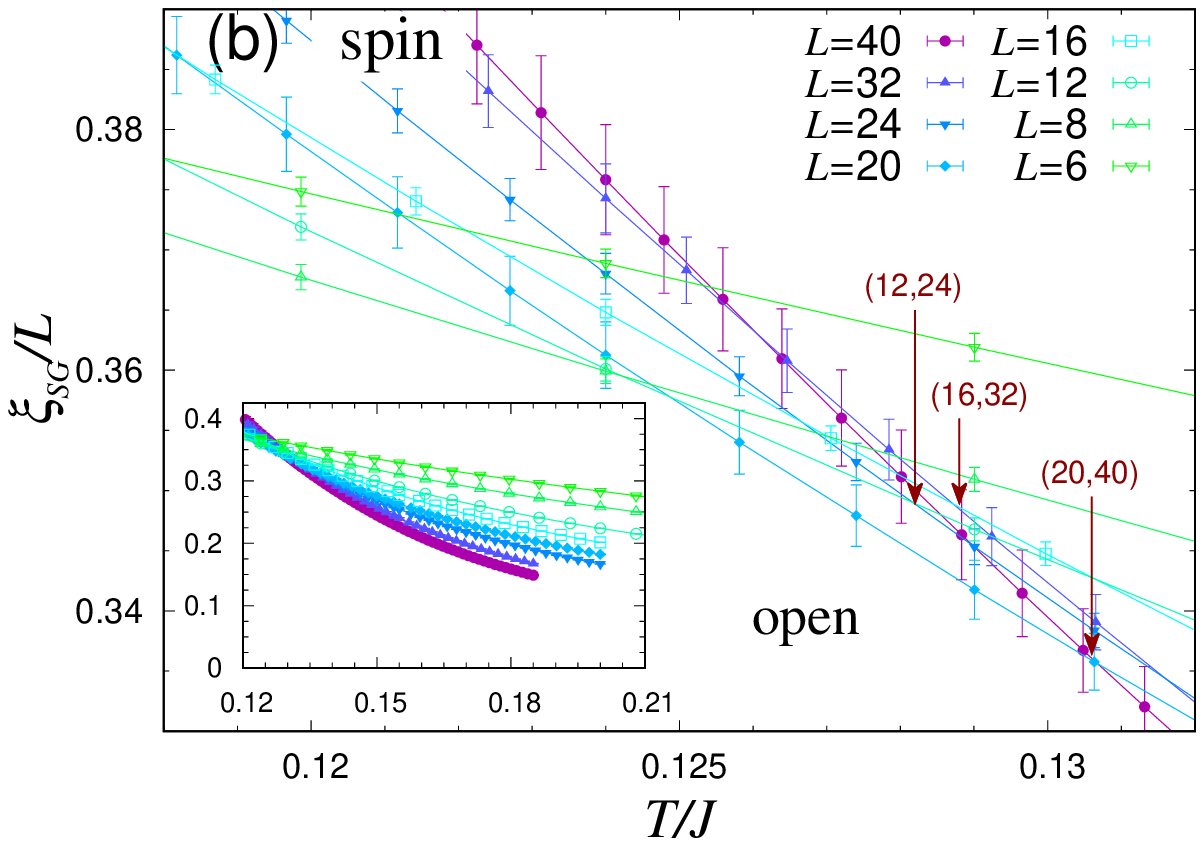}
\end{center}
\caption{
(Color online) The temperature and size dependence of the correlation-length ratio under open BC, (a) for the chirality, and (b) for the spin.  Main panel is a magnified view of the transition region, while the inset represents a wider temperature region.
}
\end{figure}

 In both Figs.1 and 2, comparison of Fig.(a) for the chirality and Fig.(b) for the spin reveals that the temperature range where the crossing occurs differs somewhat between the chirality and the spin, the former being higher than the latter (note the difference in the covered temperature range between the main figures of (a) and (b)). Of course, one needs to perform an appropriate extrapolation to the $L\rightarrow \infty$ limit to locate the bulk CG and SG transition temperatures $T_{{\rm CG}}$ and $T_{{\rm SG}}$. In the next subsection, we try to determine whether $T_{{\rm CG}}>T_{{\rm SG}}$ or $T_{{\rm CG}}=T_{{\rm SG}}$ by using the data of $T_{{\rm cross}}$ of both periodic and open BC.  

 Comparison of Fig.1 for periodic BC and Fig.2 for open BC reveals that the crossing temperatures $T_{{\rm cross}}$ for open BC tends to be lower somewhat than $T_{{\rm cross}}$ for periodic BC. Since the missing bonds at the surface tend to reduce the energy scale, the observed tendency seems to be rather natural as a surface effect. In the thermodynamic limit, however, since the bulk transition temperature should not depend on the applied BC, extrapolations of $T_{{\rm cross}}(L)$ for periodic and open BC should converge to a common value, the bulk transition temperature. We fully utilize this fact in our extrapolation procedure in the next subsection.

\begin{figure}[ht]
\begin{center}
\includegraphics[width=\hsize]{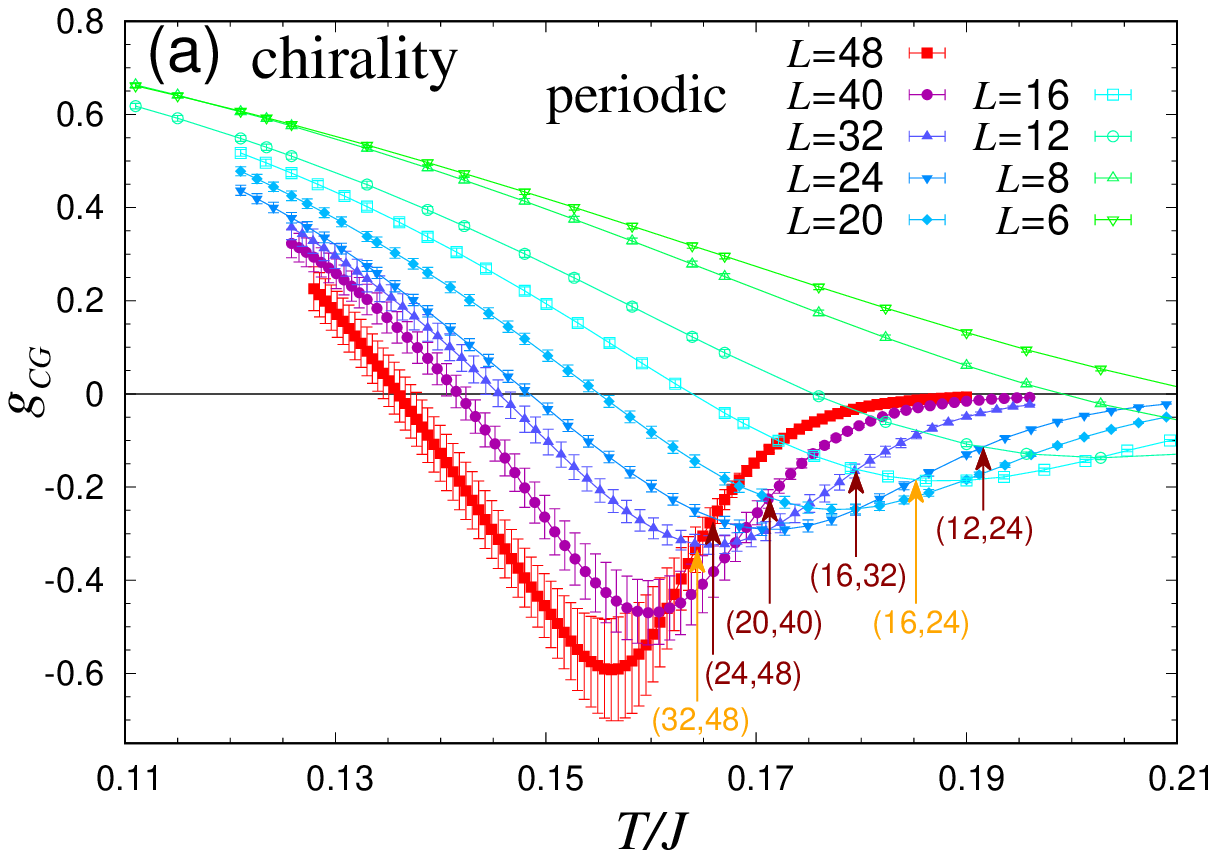}
\includegraphics[width=\hsize]{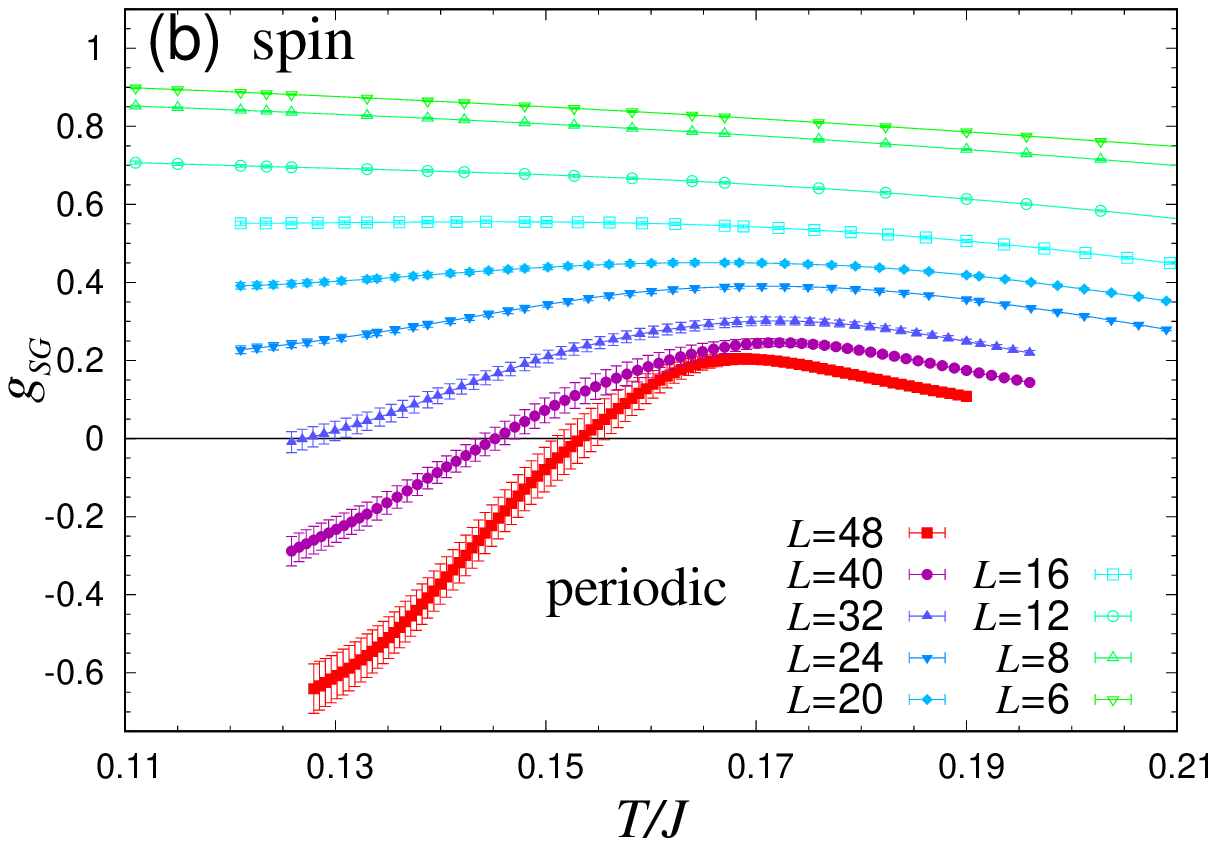}
\end{center}
\caption{
(Color online) The temperature and size dependence of the Binder ratio under periodic BC, (a) for the chirality, and (b) for the spin. 
}
\end{figure}
\begin{figure}[ht]
\begin{center}
\includegraphics[width=\hsize]{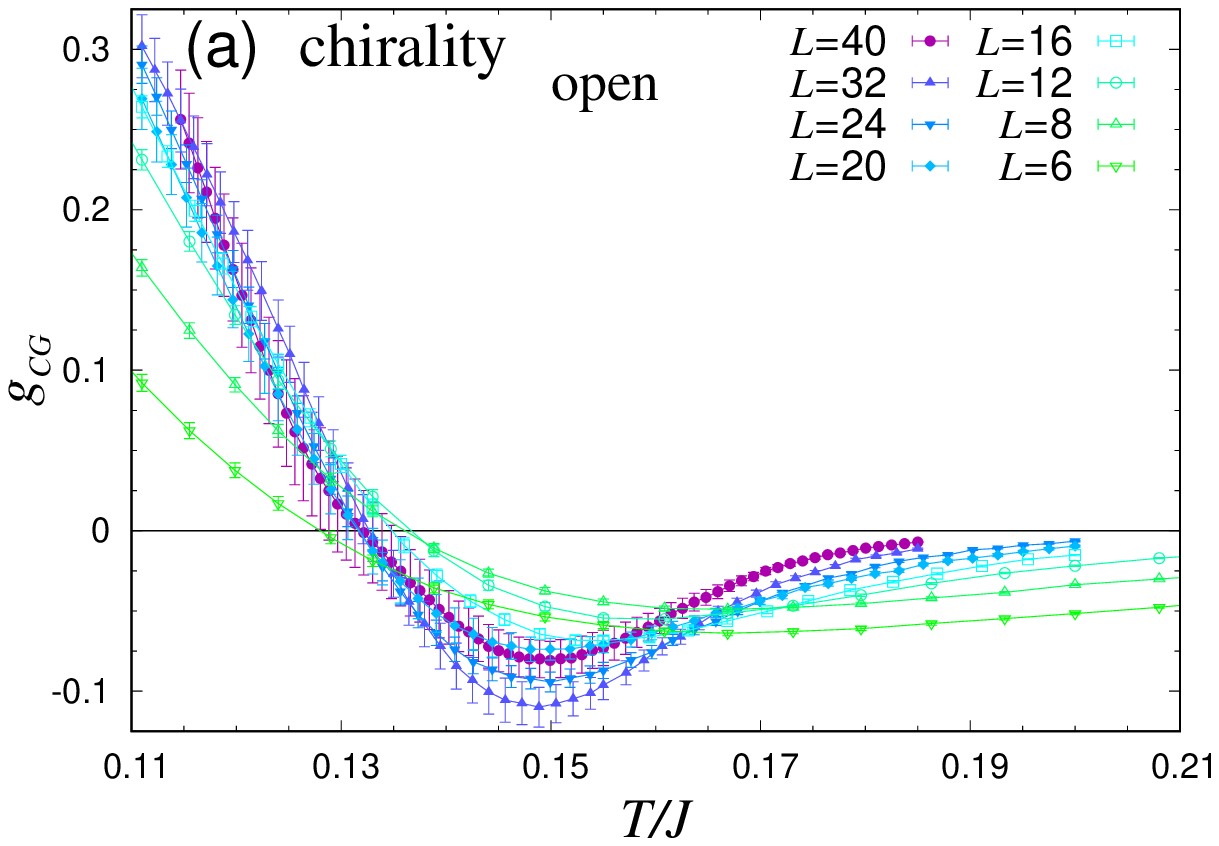}
\includegraphics[width=\hsize]{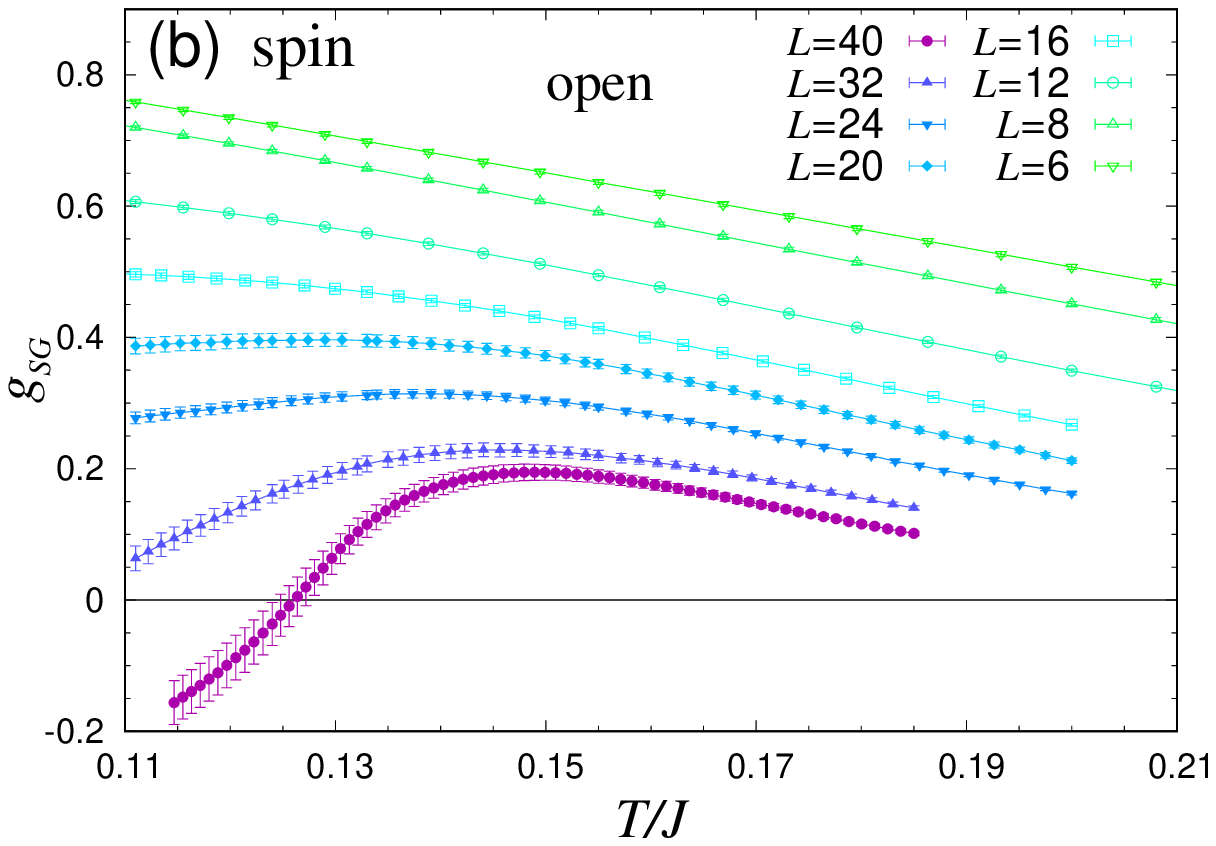}
\end{center}
\caption{
(Color online) The temperature and size dependence of the Binder ratio under open BC, (a) for the chirality, and (b) for the spin.
}
\end{figure}

 In Fig.3, the temperature and size dependence of the Binder ratio is given (a) for the chirality, and (b) for the spin, in the case of periodic BC. The corresponding figures for open BC are given in Figs.4(a) and (b), respectively. For both periodic and open BC, the CG Binder ratio $g_{{\rm CG}}$ exhibits a crossing behavior between different $L$ on the {\it negative\/} side of $g_{{\rm CG}}$, accompanied with a negative dip. This negative dip tends to deepen with increasing $L$ up to $L=48$ for periodic BC, while the tendency seems not so systematic for open BC as the dip depth becomes shallower for our largest size $L=40$. As a consequence, the crossing of $g_{{\rm CG}}$ disappears for the largest size $L=40$ for open BC. Anyway, the crossing and the dip behavior of $g_{{\rm CG}}$ is consistent with a finite-temperature transition in the chiral sector. The crossing temperature $T_{{\rm cross}}$ and the dip temperature $T_{{\rm dip}}$ should converge to the bulk $T_{{\rm CG}}$ in the thermodynamic limit as long as they persist, and can be used to locate $T_{{\rm CG}}$. We emphasize that the Binder ratio is a very useful quantity carrying valuable information in spite of the criticism of Ref.\cite{PixleyYoung}, as was demonstrated in Ref.\cite{VietKawamuraPRB}.

 As can clearly be seen from Figs.3(b) and 4(b), the SG Binder ratio $g_{{\rm SG}}$ does not exhibit any crossing, at least in the range of sizes studied, although its temperature and size dependence is quite non-trivial. In particular, $g_{{\rm SG}}$ gets negative at lower temperatures for larger sizes. Since $g_{{\rm SG}}$ is expected to converge to unity at $T=0$ in the $L\rightarrow \infty$ limit, $g_{{\rm SG}}$ would exhibit a negative dip as in the case of the CG Binder ratio $g_{{\rm CG}}$, but at a temperature much lower than the dip temperature of $g_{{\rm CG}}$, still lower than the temperature range covered by our MC simulation. The dip temperature $T_{{\rm dip}}$ of $g_{{\rm SG}}$ would converge to $T_{{\rm SG}}$ in the $L\rightarrow \infty$ limit, and if we could identify $T_{{\rm dip}}$ of $g_{{\rm SG}}$, this information would be utilized in locating $T_{{\rm SG}}$. Unfortunately, $T_{{\rm dip}}$ of $g_{{\rm SG}}$ is too low outside the investigated temperature range for the sizes studied here so that we cannot utilize $T_{{\rm dip}}$ of $g_{{\rm SG}}$. Anyway, the observation that $T_{{\rm dip}}$ of $g_{{\rm SG}}$ is significantly lower than $T_{{\rm dip}}$ of $g_{{\rm CG}}$ seems to favor the spin-chirality decoupling.

\subsection{C. The determination of $T_{{\rm CG}}$}

 In this and following subsections, on the basis of our MC data reported in the previous subsection, we try to determine the CG and SG transition temperatures as accurately as possible to examine whether the spin-chirality really occurs in the model or not. In the following, we estimate $T_{{\rm CG}}$ and $T_{{\rm SG}}$ separately by performing the $L\rightarrow \infty$ extrapolation.
 
 Generally, the crossing temperature $T_{{\rm cross}}$ between the data for the two different sizes $L$ and $sL$ ($s>1$) is expected to converge to the bulk transition temperature $T_g$ in the $L\rightarrow \infty$ limit as
\begin{equation}
T_{{\rm cross}} \approx T_g + c_s L^{-\theta}
 = T_g + c'_s L_{{\rm av}}^{-\theta},\ \ \ \theta=\frac{1}{\nu}+\omega, 
\label{Tcross}
\end{equation}
where $\nu$ is the correlation-length exponent, $\omega$ the correction-to-scaling exponent, $c'_s=c_s \left( \frac{1+s}{2} \right)^\theta$ is an $s$-dependent nonuniversal constant, and $L_{{\rm av}}=\frac{1+s}{2}L$ is the mean of the two sizes yielding the data crossing. By contrast, the dip temperature $T_{{\rm dip}}$ is expected to behave as
\begin{equation}
T_{{\rm dip}} \approx T_g + c''L^{-\frac{1}{\nu}},
\label{Tdip}
\end{equation}
where $c''$ is a nonuniversal constant.

\begin{figure}[ht]
\begin{center}
\includegraphics[width=\hsize]{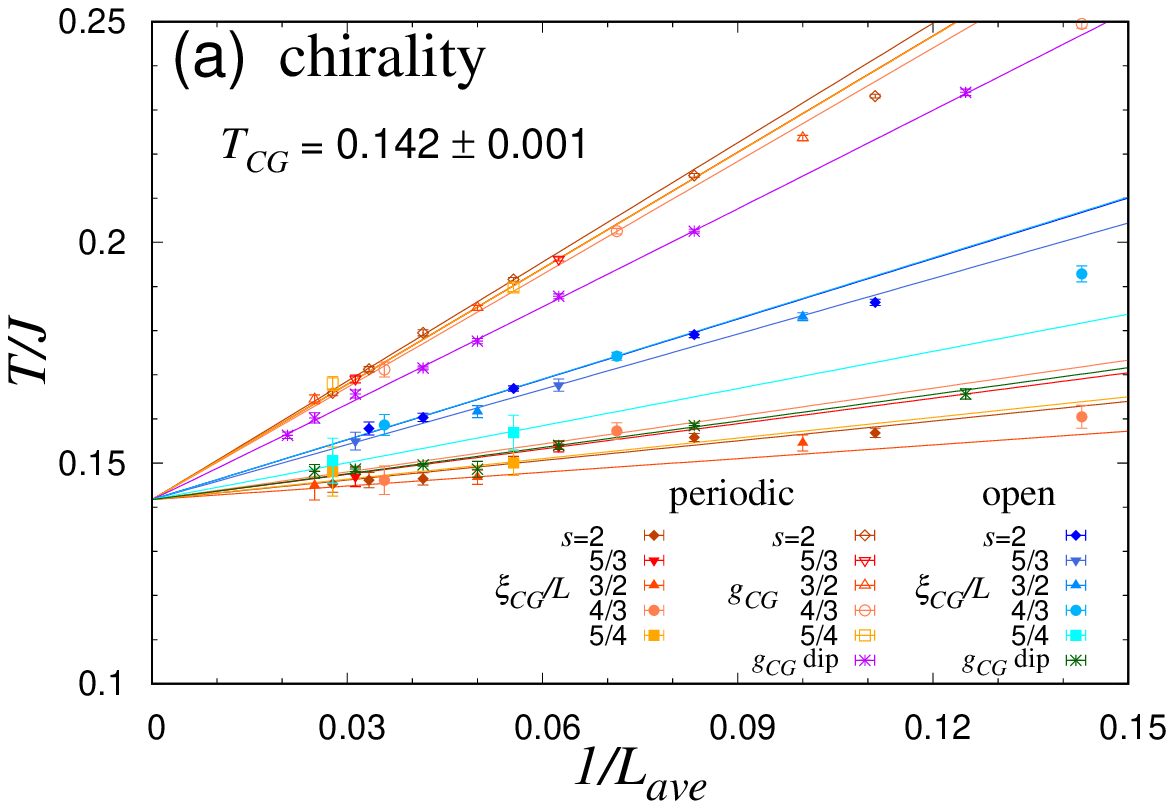}
\includegraphics[width=\hsize]{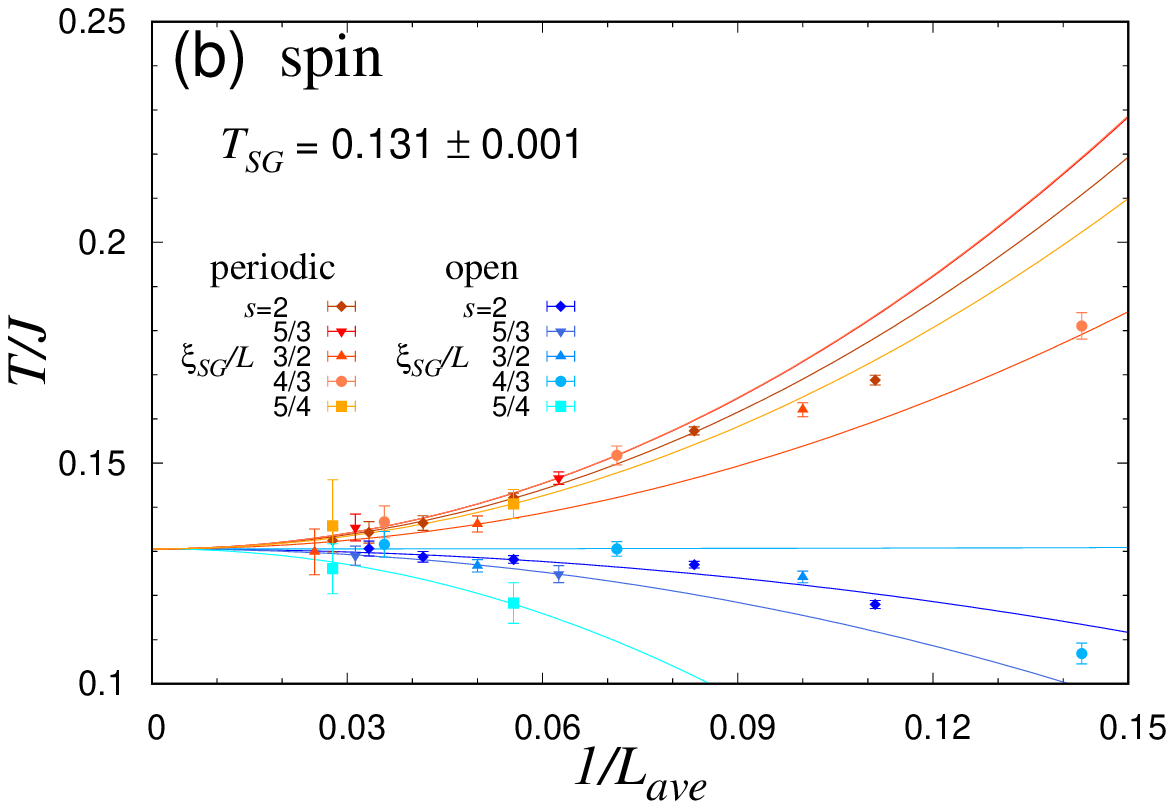}
\end{center}
\caption{
(Color online) Estimates of the CG and SG transition temperatures $T_{{\rm CG}}$ and $T_{{\rm SG}}$ on the basis of the $L\rightarrow \infty$ extrapolation of the crossing temperature $T_{{\rm cross}}$ and the dip temperature $T_{{\rm dip}}$. The crossing temperatures are between the two lattice sizes $L$ and $sL$, with $L_{{\rm av}}=\frac{1+s}{2}L$. The fits are based on Eqs.(\ref{Tcross}) and (\ref{Tdip}) by using $L\geq 12$ data ($L_{{\rm min}}=12$). (a) $T_{{\rm cross}}$ of the CG correlation-length ratio $\xi_{{\rm CG}}/L$ and $T_{{\rm dip}}$ of $g_{{\rm CG}}$ under both periodic and open BC, and $T_{{\rm cross}}$ of the CG Binder ratio $g_{{\rm CG}}$ under periodic BC, are plotted versus the inverse lattice size $1/L_{{\rm av}}$. (b) $T_{{\rm cross}}$ of the SG correlation-length ratio $\xi_{{\rm SG}}/ L$ both under periodic and open BC are plotted versus $1/L_{{\rm av}}$. The extrapolated CG and SG transition temperatures are $T_{{\rm CG}}=0.142\pm 0.001$ from (a), and $T_{{\rm SG}}=0.131\pm 0.001$ from (b), respectively.
}
\end{figure}

 In determining $T_{{\rm CG}}$, we employ the data of $T_{{\rm cross}}$ of the CG correlation-length ratio $\xi_{{\rm CG}} /L$ under both periodic and open BC, abbreviated as $T_{{\rm cross}}$[$\xi_{{\rm CG}} /L$, P] and $T_{{\rm cross}}$[$\xi_{{\rm CG}} /L$, O], $T_{{\rm cross}}$ of $g_{{\rm CG}}$ under periodic BC, $T_{{\rm cross}}$[$g_{{\rm CG}}$, P] ($T_{{\rm cross}}$ of $g_{{\rm CG}}$ under open BC is not used due to the absence of the crossing point for the $L=40$ related data), and the dip temperature of $g_{{\rm CG}}$ under both periodic and open BC, $T_{{\rm dip}}$[P] and $T_{{\rm dip}}$[O]. These data points are plotted in Fig.5 as a function of the inverse system size $1/L_{{\rm av}}$ (for $T_{{\rm dip}}$ of $g_{{\rm CG}}$, we put $L_{{\rm av}}=L$). The $L\rightarrow \infty$ extrapolation is made on the basis of Eq.(\ref{Tcross}) for $T_{{\rm cross}}$, and  on the basis of Eq.(\ref{Tdip}) for $T_{{\rm dip}}$. In the extrapolation of $T_{{\rm cross}}$, we use five distinct $s$-series of $s=2,\ \frac{5}{3},\ \frac{3}{2},\ \frac{4}{3}$ and $\frac{5}{4}$. More specifically, $s=2$ series contains (6,12), (8,16), (12,24), (16,32), (20,40) and (24,48), $s=\frac{5}{3}$ contains (12,20) and (24,40), $s=\frac{3}{2}$ contains (8,12), (16,24) and (32,48), $s=\frac{4}{3}$ contains (6,8), (12,16) and (24,32), and $s=\frac{5}{4}$ contains (16,20) and (32,40). We then perform the combined fit of $T_{{\rm cross}}$[$\xi_{{\rm CG}} /L$, P], $T_{{\rm cross}}$[$\xi_{{\rm CG}} /L$, O], $T_{{\rm cross}}$[$g_{{\rm CG}}$, P], each with $s=2,\ \frac{5}{3},\ \frac{3}{2},\ \frac{4}{3}$ and $\frac{5}{4}$, $T_{{\rm dip}}$[P] and $T_{{\rm dip}}$[O]. The fitting parameters are $T_{{\rm CG}}$, $\theta$, $\nu$, $c'_s$ for $\xi_{{\rm CG}} /L$ ($s=2,\ \frac{5}{3},\ \frac{3}{2},\ \frac{4}{3},\ \frac{5}{4}$), $c'_s$ for $g_{{\rm CG}}$ ($s=2,\ \frac{5}{3},\ \frac{3}{2},\ \frac{4}{3},\ \frac{5}{4}$) and $c''$.

 If we use all sizes of $L\geq 6$ in the combined fit, there are 63 data points (16 from $T_{{\rm cross}}$[$\xi_{{\rm CG}} /L$, P], 16 from $T_{{\rm cross}}$[$g_{{\rm CG}} /L$, P], 14 from $T_{{\rm cross}}$[$\xi_{{\rm CG}} /L$, O], 9 from $T_{{\rm dip}}$[P], 8 from  $T_{{\rm dip}}$[O]), and 20 fitting parameters to be determined, enough number of degrees of freedom (NDF=63-20=43) being left in the fit. The fit yields $T_{{\rm CG}}=0.142\pm 0.001$ with the reduced $\chi^2$-value of 4.57. Since we assume only the leading scaling form in Eqs.(\ref{Tcross}) and (\ref{Tdip}), too small sizes might lie outside the critical regime and better ruled out from the fit. In order to examine the possible effect of finite sizes systematically, we also try the similar combined fits with varying the minimum lattice size used in the fit, $L_{{\rm min}}$, increasing $L_{{\rm min}}$ from 6 to larger values. Yet, even if we choose $L_{{\rm min}}=12$ or 16, we still obtain $T_{{\rm CG}}=0.142\pm 0.001$ with the NDF=27, $\chi^2=0.76$ ($L_{{\rm min}}=12$), and $T_{{\rm CG}}=0.142\pm 0.001$ with the NDF=16, $\chi^2=0.83$ ($L_{{\rm min}}=16$). The $\chi^2$-value becomes minimum for $L_{{\rm min}}=12$, and the $\chi^2$-value itself is quite reasonable. We then regard the $L_{{\rm min}}=12$ as optimal, and the corresponding fit with $L_{{\rm min}}=12$ is shown in Fig.5. We note that even if we cut the the larger-size data in the fit, {\it e.g.\/}, cut $L=48$, the fit still yields almost identical estimate of $T_{{\rm CG}}=0.142\pm 0.001$. Thus, our estimate of $T_{{\rm CG}}$ is quite stable, insensitive to the $L$-range considered. In particular, there is no indication of the size crossover occurring up to the size $L=48$.

 The earlier large-scale MC simulations on the model were made under periodic BC \cite{LeeYoung2003,Campos,LeeYoung2007,VietKawamuraPRL,VietKawamuraPRB,Fernandez}, and the transition temperature was estimated based only on the periodic-BC data. In contrast, our present simulation and analysis are made both under periodic and open BC. In order to see how this affects the estimate of $T_{{\rm CG}}$, we also perform the similar analysis by using only the periodic-BC data. Of course, the number of data points and the NDF are reduced. Yet, the resulting estimate of $T_{{\rm CG}}$ has tuned out to remain almost the same: for $L_{{\rm min}}=12$, our estimate is $T_{{\rm CG}}=0.141\pm 0.002$ with NDF=17 and $\chi^2=0.76$, quite consistent with our estimate above based on both periodic and open BC. In view of these results, we finally quote $T_{{\rm CG}}=0.142\pm 0.001$.

 This present estimate of $T_{{\rm CG}}$ is fully consistent with the earlier estimate of Ref.\cite{VietKawamuraPRB}, $T_{{\rm CG}}=0.143\pm 0.003$, but largely deviate from that of Ref.\cite{Fernandez}, $T_{{\rm CG}}=T_{{\rm SG}}=0.120^{+0.010}_{-0.100}$. We shall discuss the possible cause of this deviation later in the next subsection.

\subsection{D. The determination of $T_{{\rm SG}}$}

 We now move on to the estimate of the SG transition temperature $T_{{\rm SG}}$. As mentioned, the SG Binder ratio does not exhibit any crossing nor a dip in the range of sizes studied so that the data available to estimate $T_{{\rm SG}}$ comes only from $T_{{\rm cross}}$ of $\xi_{{\rm SG}} /L$, {\it i.e.\/}, $T_{{\rm cross}}$[$\xi_{{\rm SG}} /L$, P] and $T_{{\rm cross}}$[$\xi_{{\rm SG}} /L$, O]. These $T_{{\rm cross}}$ data are shown in Fig.5(b). Interestingly, $T_{{\rm cross}}$ exhibits opposite size dependence between the two types of BC. Namely, with increasing $L$, $T_{{\rm cross}}$ tends to decrease for periodic BC, while it tends to {\it increase\/} for open BC, almost saturating for larger sizes studied.

 Based on these data of $T_{{\rm cross}}$, we perform the fit on the basis of Eq.(\ref{Tcross}). The available $s$-series is the same as in the CG case, {\it i.e.\/}, $s=2,\ \frac{5}{3},\ \frac{3}{2},\ \frac{4}{3},\ \frac{5}{4}$. If we use all the sizes in the fit, {\it i.e.\/}, if $L_{{\rm min}}=6$, the combined fit yields $T_{{\rm SG}}=0.132\pm 0.001$ with NDF=18 and $\chi^2=3.08$. For $L_{{\rm min}}=12$, the fit yields $T_{{\rm SG}}=0.131\pm 0.001$ with NDF=10 and $\chi^2=0.13$, while for $L_{{\rm min}}=16$, $T_{{\rm SG}}=0.131\pm 0.001$ with NDF=4 and $\chi^2=0.19$. Thus, the fit turns out to be fairly stable also for $T_{{\rm SG}}$.

 We also perform the similar analysis by using only the periodic-BC data, although the NDF becomes even smaller. For $L_{{\rm min}}=12$, our estimate is $T_{{\rm SG}}=0.126\pm 0.003$ with NDF=5 and $\chi^2=0.08$ (the $\chi^2$-value seems a bit too small). This estimate is quite close to the earlier estimate of Ref.\cite{VietKawamuraPRB} based on the periodic-BC data, $T_{{\rm SG}}=0.125^{+0.006}_{-0.0.012}$, but lower somewhat from our present estimate based on both periodic and open BC, $T_{{\rm SG}}=0.131\pm 0.001$. Indeed, the open-BC data approach $T_{{\rm SG}}$ from below in the range of sizes studied, leading to the higher present estimate of $T_{{\rm SG}}$ as can be seen from Fig.5(b). If this increasing and saturating trend of $T_{{\rm cross}}$ vs. $1/L_{{\rm av}}$ persists for still larger $L$, somewhat higher estimate of $T_{{\rm SG}}\simeq 0.131$ would be justified. By contrast, if the increasing and saturating trend of $T_{{\rm cross}}$ vs. $1/L_{{\rm av}}$ would exhibit a ``turnover'' to the decreasing trend for still larger lattices, the lower estimate of $T_{{\rm SG}}\simeq 0.126$ might well be reasonable. In fact, we cannot rule out such a possibility from our present data. In view of all these, we finally quote $T_{{\rm SG}}=0.131^{+0.001}_{-0.006}$. This estimate of $T_{{\rm SG}}$ is consistent both with the ones reported in Ref.\cite{VietKawamuraPRB} $T_{{\rm SG}}=0.125^{+0.006}_{-0.012}$, and in Ref.\cite{Fernandez} $T_{{\rm SG}}=0.129^{+0.003}_{-0.016}$.

 Our final estimate of the transition temperatures are then
\begin{equation}
T_{{\rm CG}}=0.142\pm 0.001,\ \ T_{{\rm SG}}=0.131^{+0.001}_{-0.006}.
\end{equation}
The result strongly supports the occurrence of the spin-chirality decoupling in the model.

 If we compare these estimates with the relatively recent large-scale MC simulations \cite{VietKawamuraPRL,VietKawamuraPRB,Fernandez}, the only sizable difference arises from the $T_{{\rm CG}}$-value reported in Ref.\cite{Fernandez}, $T_{{\rm CG}}(=T_{{\rm SG}})=0.120^{+0.010}_{-0.100}$. We wish to discuss the cause of this discrepancy. Fernandez {\it et al\/} simulated the sizes of $L=8$,12, 16, 24, 32 and 48 under periodic BC only (the maximum size is the same as our present one) and employed the crossing temperatures of the CG correlation-length ratio with the $s$-series only of $s=2$ and 3/2. Setting $L_{{\rm min}}=12$ and applying the fitting form of Eq.(\ref{Tcross}) to the $T_{{\rm cross}}$[$\xi_{{\rm CG}} /L$, P] data with $s=2$ and 3/2, Fernandez {\it et al\/} obtained a {\it negative\/} $T_{{\rm CG}}$. A negative $T_{{\rm CG}}$ is clearly problematic because there is now consensus that the 3D Heisenberg SG exhibits a finite-temperature transition with the noncoplanar order characterized by the nonzero chirality, even if the presence of the spin-chirality decoupling might still be at issue. Then, these authors {\it forced\/} the relation $T_{{\rm CG}}=T_{{\rm SG}}$ by assuming the absence of the spin-chirality decoupling, and by combining their CG and SG correlation-length ratios data with a common $T_g$, obtained an estimate quoted above. % We note that the amount of information they employed in estimating $T_{{\rm CG}}$ is rather limited.

 Since our present data yield quite stable estimate of $T_{{\rm CG}}=0.142\pm 0.001$ as detailed above, an unphysical negative $T_{{\rm CG}}$ reported in Ref.\cite{Fernandez} looks strange to us, and we try to further clarify the situation.  If we repeat the same fit by using the $T_{{\rm cross}}$[$\xi_{{\rm CG}} /L$, P] data reported in Ref.\cite{Fernandez} with $L_{{\rm min}}=12$, we get $T_{{\rm CG}}=-4\pm 600$ with the NDF=1 and $\chi^2=0.16$. As reported in Ref.\cite{Fernandez}, a negative $T_{{\rm CG}}$-value certainly comes out, but the associated error bar is unusually large, and nothing can actually be concluded concerning the presence or absence of the spin-chirality decoupling. To check whether such a large error bar is originated just from the small NDF(=1), we try the same type of fit by using our own data of $T_{{\rm cross}}$[$\xi_{{\rm CG}} /L$, P] with $L_{{\rm min}}=12$, to obtain $T_{{\rm CG}}=0.145\pm 0.001$, which is positive and turns out to be close to the $T_{{\rm CG}}$-value obtained from our full analysis. We find that such a large difference in the estimates of $T_{{\rm CG}}$ comes, not just from the difference in the NDF, but also from the difference in the $T_{{\rm cross}}$-values related to the largest size $L=48$, {\it i.e.\/}, $T_{{\rm cross}}=0.142(1)$ for (24,48) ($s=2$) in Ref.\cite{Fernandez} vs. 0.145(2) in our present computation, while 0.138(3) for (32,48) ($s=\frac{3}{2}$) in Ref.\cite{Fernandez} vs. 0.145(3) in our present computation. Hence, the $L=48$-related $T_{{\rm cross}}$ of Ref.\cite{Fernandez} is lower than our present values by $2\sim 3\sigma$. This difference, which may not necessarily be a contradiction in the numerical sense, is combined with the small NDF(=1), and is causing a negative $T_{{\rm CG}}$ estimate with quite a large error bar mentioned above, and eventually the large difference in the final estimate of $T_{{\rm CG}}$.

 In any case, in view of the stability of our estimate of $T_{{\rm CG}}$ and $T_{{\rm SG}}$ against not only the lattice sizes but also several distinct types of independent physical quantities and BC, we believe that our present estimates of $T_{{\rm CG}}$ and $T_{{\rm SG}}$ are trustable, and presents a strong numerical evidence of the spin-chirality decoupling.

\subsection{E. The $\xi_{{\rm CG}}$ versus $\xi_{{\rm SG}}$ relation}

 In the previous subsections, we have established that $T_{{\rm CG}} > T_{{\rm SG}}$, {\it i.e.\/}, the spin-chirality decoupling. This means, in the thermodynamic limit $L\rightarrow \infty$, the CG correlation length $\xi_{{\rm CG},\infty}$ outgrows the SG correlation length $\xi_{{\rm SG},\infty}$ slightly above $T_{{\rm CG}}$, $\xi_{{\rm CG,\infty}} > \xi_{{\rm SG,\infty}}$. For finite size system, this inequality does not necessarily hold. In fact, for smaller sizes $L$, the opposite inequality $\xi_{{\rm CG},L} < \xi_{{\rm SG},L}$ usually holds. This is quite natural since at shorter length scale the chirality is a composite of spins, the chiral order being parastic to the spin order. Indeed, Viet and Kawamura observed in Ref.\cite{VietKawamuraPRB} that the inequality $\xi_{{\rm CG},L} < \xi_{{\rm SG},L}$ always held in the investigated temperature range for $L\leq 32$, although the ratio $\xi_{{\rm CG},L}/\xi_{{\rm SG},L}$ monotonically increased with increasing $L$ toward unity. In our present calculation, we simulate larger lattices than those in Ref.\cite{VietKawamuraPRB}, $L=40$  and 48, and it might be interesting to plot the ratio $\xi_{{\rm CG},L}/\xi_{{\rm SG},L}$ in the same way as in Ref.\cite{VietKawamuraPRB}. The resulting figure under periodic BC is given in Fig.6. One can see from the figure that the ratio $\xi_{{\rm CG},L}/\xi_{{\rm SG},L}$ now exceeds unity at lower temperatures for $L=40$ and 48, {\it i.e.\/}, the CG correlation length outgrows the SG correlation length. The temperature at which the ratio exceeds unity tends to increase as $L$ increases. The crossing temperature of $L=48$ already exceeds $T_{{\rm SG}}$. We emphasize that nothing special happens when the ratio exceeds unity with increasing $L$ and lowering $T$, {\it i.e.\/}, the CG correlation length simply outgrows the SG correlation length without any `hesitation', {\it i.e.\/}, no sign of merging nor saturation. The result gives a strong support for the spin-chirality decoupling really taking place in the model. We note that the similar `outgrowth' of $\xi_{{\rm CG},L}$ over $\xi_{{\rm SG},L}$ was observed in the 3D {\it XY\/} SG with two-component spins in Ref.\cite{ObuchiKawamura}, where $\xi_{{\rm CG},L}$ outgrew $\xi_{{\rm SG},L}$ at low temperatures for the largest size studied $L=40$ ($L=48$ data not available there).

\begin{figure}[ht]
\begin{center}
\includegraphics[width=\hsize]{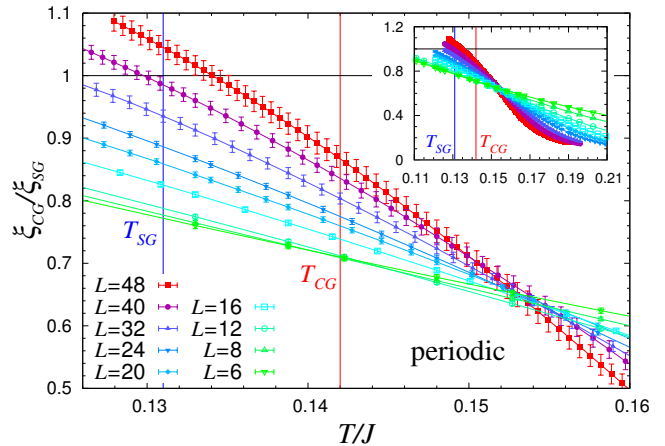}
\end{center}
\caption{
(Color online) The temperature and size dependence of the ratio of the CG correlation length and the SG correlation length, $\xi_{{\rm CG},L}/\xi_{{\rm SG},L}$, in the transition region under periodic BC. For larger sizes $L=40$ and 48, the ratio exceeds unity at low temperatures. The inset represents a wider temperature range.
}
\end{figure}

\subsection{F. The nature of the RSB in the chiral-glass state}

\begin{figure}[ht]
\begin{center}
\includegraphics[width=\hsize]{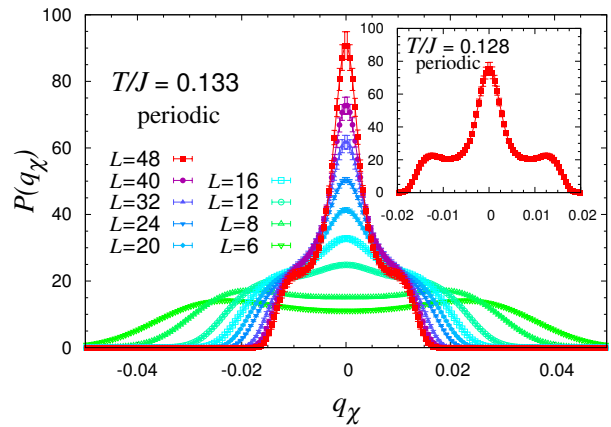}
\end{center}
\caption{
(Color online) The chiral overlap distribution function in the CG state under periodic BC at a temperature $T=0.133$ for various lattice sizes. The one at $T=0.128$ is shown in the inset for the largest size $L=48$. 
}
\end{figure}

 After establishing the existence of the spin-chirality decoupling and the CG state, we now wish to investigate the nature of the CG ordered state itself, the nature of the possible RSB, in particular. Some time ago, Hukushima and Kawamura proposed on the basis of MC simulation that the CG ordered state might exhibit a peculiar one-step-like RSB in the chiral sector \cite{HukushimaKawamura2000}, in sharp contrast to the Ising SG exhibiting the hierarchical RSB \'a la Parisi. The proposal was supported also by the later simulation on the model \cite{HukushimaKawamura2005,VietKawamuraPRB}. Interestingly, in the chirality scenario, this observation means that the SG order of real weakly anisotropic Heisenberg-like SG including canonical SG also exhibits the one-step-like RSB. Indeed, some support was already reported from off-equilibrium properties of the weakly anisotropic Heisenberg-like SG either numerically \cite{Kawamura2003} or experimentally \cite{HerissonOcio2002,HerissonOcio2004}. 

 In order to get further information on the issue, we compute the chiral-overlap distribution $P(q_{\chi})$, which is defined by 
\begin{equation}
P(q_{\chi}^{'})=[\langle \delta(q_{\chi}^{'}-q_{\chi})\rangle].
\end{equation}
The computed $P(q_{\chi})$ in the CG state under periodic BC is shown in Fig.7. In the main panel, the size dependence of $P(q_{\chi})$ is shown at a temperature $T=0.133$ lying in the CG state. There exists an eminent central $q_\chi=0$ component, which grows sharply with the system size $L$, suggestive of the one-step-like feature of the CG ordered state. In addition, symmetric side peaks or shoulders appear at $q_\chi=\pm q_{{\rm CG}}^{{\rm EA}}$ corresponding to the long-range CG order ($q_{{\rm CG}}^{{\rm EA}}$ represents the EA order parameter associated with the CG order). In the inset, $P(q_{\chi})$ for the largest size $L=48$ is shown at a lower temperature $T=0.128$ close to the SG phase boundary, where the side peaks become clearer.  Thus, our present data further strengthens the one-step-like RSB character of the CG order of the model. Such a one-step-like feature is also consistent with the observed negative Binder ratios $g_{{\rm CG}}$ and $g_{{\rm SG}}$ at lower temperatures: see Figs.3(b) and 4(b).

 It should be noticed that the computed $P(q_{\chi})$ is also consistent with a continous plateau part characteristic of the full-step RSB superimposed on the $q_\chi=0$ central peak characteristic of the one-step RSB and the $q_\chi=\pm q_{{\rm CG}}^{{\rm EA}}$ peaks. If that is the case, the RSB associated with the CG order might be the combination of the pure one-step ($q_\chi=0$ central peak) and the full-step (continuous plateau spanning between $q_\chi=\pm q_{{\rm CG}}^{{\rm EA}}$).

\section{IV. The critical properties}

In this section, on the basis of the $T_{{\rm CG}}$- and  $T_{{\rm SG}}$-values determined in the previous section, we wish to examine the critical properties associated with the CG and SG transitions. Periodic BC is better suited for this purpose, since the open-BC data are complicated by the possible contribution from the surface critical phenomena. Hence, we concentrate on the periodic-BC data in this section. We consider the effect of the correction-to-leading scaling by introducing the correction-to-scaling exponent $\omega$ in the analysis. In estimating critical exponents, we employ the Baysian scaling analysis \cite{Harada}.

\subsection{A. Chiral-glass critical properties}

We begin with the critical properties associated with the CG transition at $T=T_{{\rm CG}}$. As can clearly be seen from Fig.1(a), the crossing temperatures of the CG correlation-length ratio $\xi_{{\rm CG}} /L$ exhibit the non-negligible size dependence, indicating the necessity to consider the correction term to properly account for the CG critical properties. The appropriate scaling form with the correction term might be given by
\begin{eqnarray}
\frac {\xi_{{\rm CG}}}{L}=\tilde X_{{\rm CG}} ((T-T_{{\rm CG}})L^{\frac{1}{\nu_{{\rm CG}}}})(1+a_\xi L^{-\omega_{{\rm CG}}}),
\label{scaling-xiCG}
\end{eqnarray}
where $a_\xi$ is a numerical constant. We put $T_{{\rm CG}}=0.142$ as determined above, and try to fit the data to Eq.(\ref{scaling-xiCG}) by adjusting $\nu_{{\rm CG}}$ and $\omega_{{\rm CG}}$. The best fit is obtained for $\nu_{{\rm CG}}=1.36$ and $\omega_{{\rm CG}}=0.38$, which is shown in Fig.8(a).

\begin{figure}[ht]
\begin{center}
\includegraphics[width=\hsize]{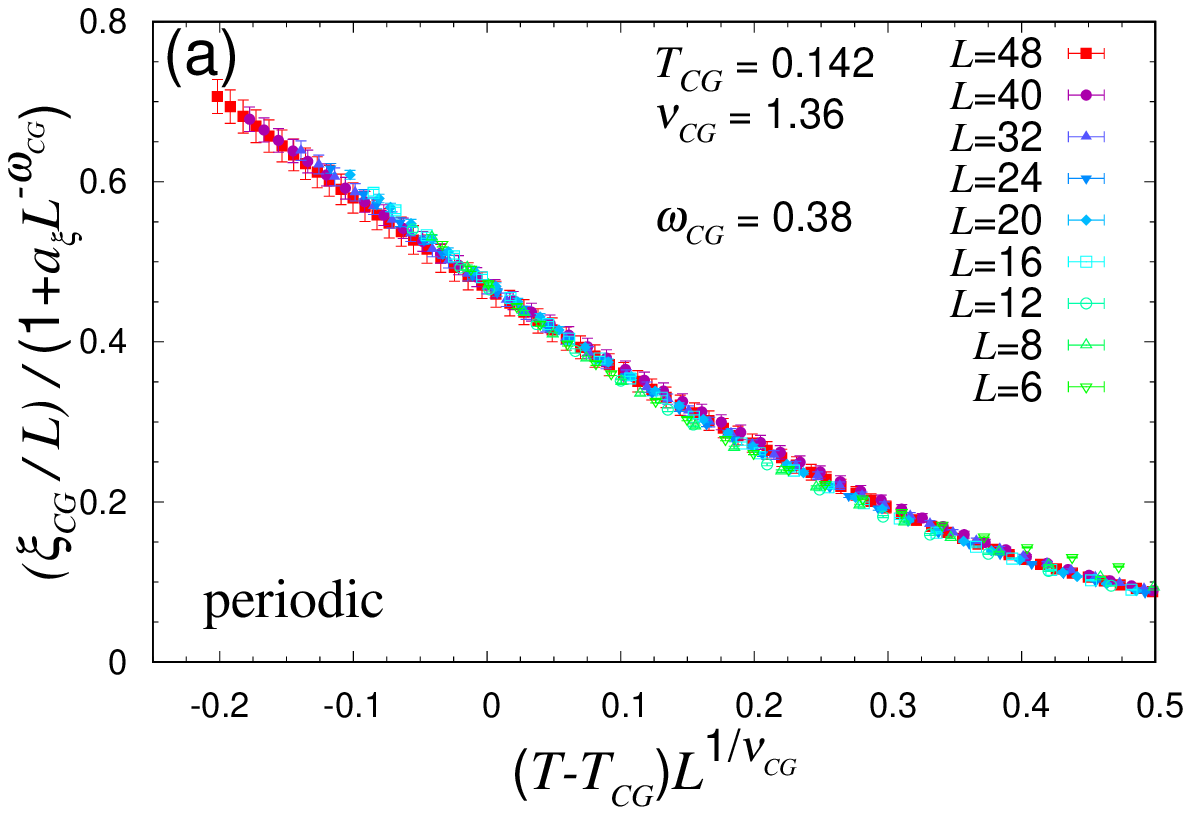}
\includegraphics[width=\hsize]{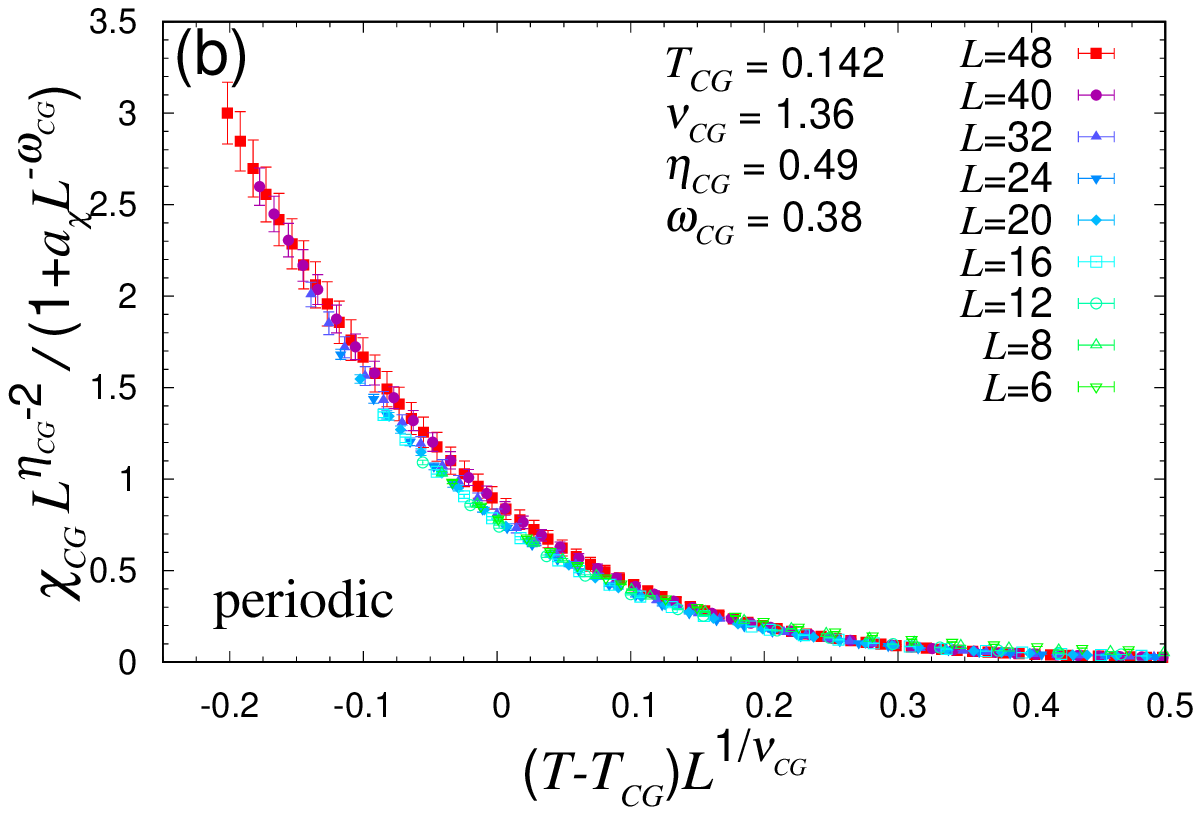}
\end{center}
\caption{
(Color online) Finite-size-scaling plots, (a) of the CG correlation-length ratio $\xi_{{\rm CG}}/L$, and (b) of the CG susceptibility $\chi_{{\rm CG}}$, under periodic BC where the correction-to-scaling effect is taken into account. The CG transition temperature is fixed to $T_{{\rm CG}}=0.142$ as determined in \S IIIC. The best fit for $\xi_{{\rm CG}}/L$ is obtained with $\nu_{{\rm CG}}=1.36$ and $\omega_{{\rm CG}}=0.38$, while that for $\chi_{{\rm CG}}$ is obtained with $\eta_{{\rm CG}}=0.49$.
}
\end{figure}

 We also try a similar finite-size scaling analysis for the CG susceptibility $\chi_{{\rm CG}}$ taking account of the correction term. The appropriate scaling form is given by
%
%\begin{eqnarray}
\begin{equation}
\chi_{{\rm CG}}=L^{2-\eta_{{\rm CG}}}\tilde Y_{{\rm CG}}((T-T_{{\rm CG}})L^{\frac{1}{\nu_{{\rm CG}}}})(1+a_\chi L^{-\omega_{{\rm CG}}}),
\label{scaling-chiCG}
%\end{eqnarray}
\end{equation}
where $a_\chi$ is a numerical constant. We put $T_{{\rm CG}}=0.142$, $\nu_{{\rm CG}}=1.36$ and $\omega_{{\rm CG}}=0.38$ as determined above, and try to fit the data to Eq.(\ref{scaling-chiCG}) by adjusting $\eta_{{\rm CG}}$. The best fit is obtained for $\eta_{{\rm CG}}=0.49$, which is shown in Fig.8(b). By examining the dependence of the resulting exponents on the choice of $L_{{\rm min}}$ and on the uncertainty of $T_{{\rm CG}}$, we estimate the error bars as
\begin{equation} 
\nu_{{\rm CG}}=1.36\pm 0.10,\ \ \eta_{{\rm CG}}=0.49 \pm 0.10,\ \ \omega_{{\rm CG}}=0.38^{+0.30}_{-0.10}.
\end{equation}
 The obtained CG exponents $\nu_{{\rm CG}}=1.36$ and $\eta_{{\rm CG}}=0.49$ are rather close to the values  previously reported in Ref.\cite{VietKawamuraPRB}, $\nu_{{\rm CG}}=1.4\pm 0.2$ and $\eta_{{\rm CG}}=0.6\pm 0.2$. With use of the scaling and hyperscaling relations, other exponents are estimated as
\begin{eqnarray} 
\alpha_{{\rm CG}}=-2.1\pm 0.3&,&\ \ \beta_{{\rm CG}}=1.0 \pm 0.1,\\ 
\gamma_{{\rm CG}}=2.1\pm 0.2&,&\ \ \delta_{{\rm CG}}=3.0 \pm 0.3.
\end{eqnarray}

\subsection{B. Spin-glass critical properties}

Next, we move to the SG critical properties. The SG transition temperature was estimated to be $T_{{\rm SG}}=0.131^{+0.001}_{-0.006}$ in the previous section. First, we set the $T_{{\rm SG}}$-value to our best estimate $T_{{\rm SG}}=0.131$, and try to scale the SG correlation-length ratio based on the scaling form,
\begin{equation}
\frac {\xi_{{\rm SG}}}{L}=\tilde X_{{\rm SG}} ((T-T_{{\rm SG}})L^{\frac{1}{\nu_{{\rm SG}}}})(1+a'_\xi L^{-\omega_{{\rm SG}}}).
\label{scaling-xiSG}
\end{equation}
In the fit, we employ the data lying in the temperature range $T<0.142=T_{{\rm CG}}$ to avoid the possible influence of the CG transition on the SG critical properties. Then, the best fit is obtained for $\nu_{{\rm SG}}=1.20$ and $\omega_{{\rm SG}}=1.32$, as shown in Fig.9(a). 

 Similar scaling analysis is made also for the SG susceptibility  based on the scaling form,
\begin{equation}
\chi_{{\rm SG}}=L^{2-\eta_{{\rm SG}}}\tilde Y_{{\rm SG}}((T-T_{{\rm SG}})L^{\frac{1}{\nu_{{\rm SG}}}})(1+a'_\chi L^{-\omega_{{\rm SG}}}).
\label{scaling-chiSG}
\end{equation}
By fixing $\nu_{{\rm SG}}=1.20$ and $\omega_{{\rm SG}}=1.32$ as determined above, $\eta_{{\rm SG}}$ is determined to be $\eta_{{\rm SG}}=-0.22$, and the corresponding best fit is given in Fig.9(b). Note that, unlike $\eta_{{\rm CG}}$, $\eta_{{\rm SG}}$ takes a negative value, whereas $\nu_{{\rm SG}}\simeq 1.20$ is not so different from $\nu_{{\rm CG}}\simeq 1.36$.
\begin{figure}[ht]
\begin{center}
\includegraphics[width=\hsize]{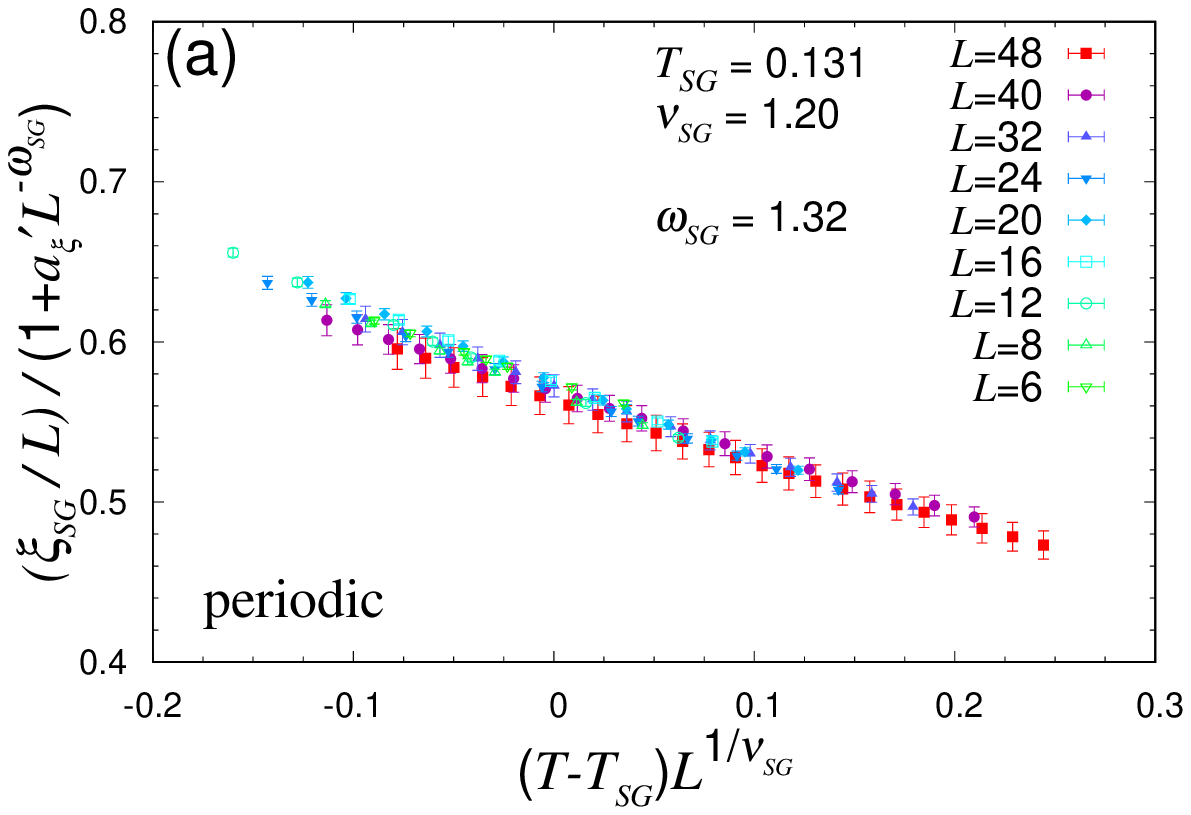}
\includegraphics[width=\hsize]{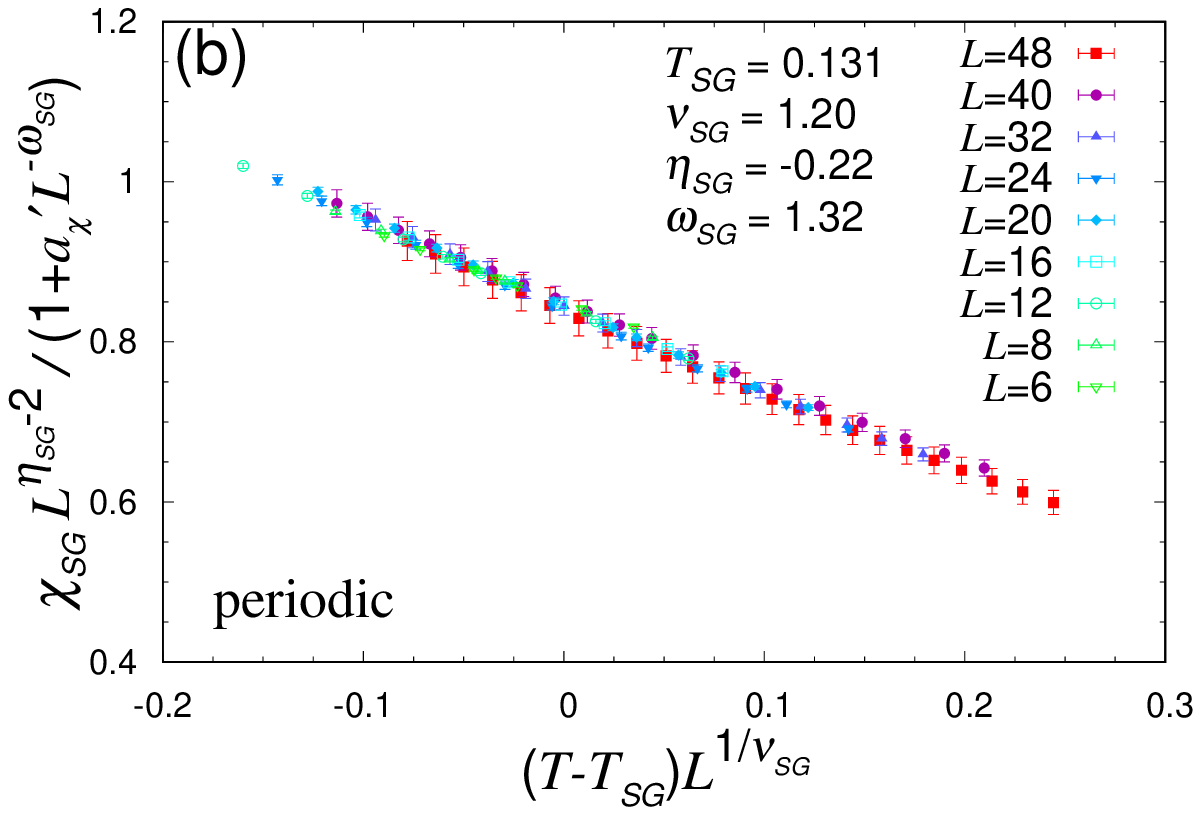}
\end{center}
\caption{
(Color online) Finite-size-scaling plots, (a) of the SG correlation-length ratio $\xi_{{\rm SG}}/L$, and (b) of the SG susceptibility $\chi_{{\rm SG}}$, under periodic BC where the correction-to-scaling effect is taken into account. The SG transition temperature is fixed to $T_{{\rm SG}}=0.131$ as determined in \S IIID. The best fit for $\xi_{{\rm SG}}/L$ is obtained with $\nu_{{\rm SG}}=1.20$ and $\omega_{{\rm SG}}=1.32$, while that for $\chi_{{\rm SG}}$ is obtained with $\eta_{{\rm SG}}=-0.22$.
}
\end{figure}

 As detailed in the previous section, our estimate of $T_{{\rm SG}}$ has a larger error bar on the lower-temperature side because the use of only periodic-BC data yields a lower estimate of $T_{{\rm SG}}\simeq 0.126$ . In view of this uncertainty, we also examine the finite-size scaling by assuming $T_{{\rm SG}}=0.126$. From the scaling of $\xi_{{\rm SG}} /L$, we get $\nu_{{\rm SG}}=1.19$ and $\omega_{{\rm SG}}=0.75$, and $\eta_{{\rm SG}}=-0.31$ from the scaling of $\chi_{{\rm SG}}$.

 In view of these observations, and also examining the dependence on the choice of $L_{{\rm min}}$, we finally quote as the SG critical exponents
\begin{equation}
\nu_{{\rm SG}}=1.2\pm 0.1,\ \  \eta_{{\rm SG}}=-0.25\pm 0.10,\ \ \omega_{{\rm SG}}=1.32\pm 0.40.
\end{equation}
 With use of the scaling and hyperscaling relations, other exponents are estimated as
\begin{eqnarray} 
\alpha_{{\rm SG}}=-1.6\pm 0.3&,&\ \ \beta_{{\rm SG}}=0.45 \pm 0.05,\\
\gamma_{{\rm SG}}=2.7\pm 0.3&,&\ \ \delta_{{\rm SG}}=7.0 \pm 1.1.
\end{eqnarray}
If we compare these SG exponents with the corresponding CG exponents determined above, $\nu$ and $\alpha$ are rather close, while $\eta$, $\beta$ and $\delta$ are significantly different. For $\eta$, the sign is opposite. For $\beta$, the SG $\beta$ is about half the CG $\beta$, whereas for $\delta$ the SG $\delta$ is about twice the CG $\delta$.

\section{V. Summary and discussion}

\subsection{Summary of the present results}

 In the present paper, we performed an extensive MC simulation on the 3D isotropic Heisenberg SG model with the random nearest-neighbor Gaussian coupling, the same model as studied previously by many authors. Our maximum size was $L=48$, the same as the largest size studied before, but we simulated both periodic BC and open BC in parallel, and utilized the both data in locating the transition temperatures $T_{{\rm CG}}$ and $T_{{\rm SG}}$. In addition, we computed and utilized a variety of independent physical quantities, not only the crossing temperatures of the correlation-length ratio under periodic BC utilized in Refs.\cite{Campos,LeeYoung2003,LeeYoung2007,Fernandez}, but also those under open BC, and the crossing temperatures and the dip temperatures of the Binder ratio as well. Our strategy was to utilize as many independent information (data points) as possible to get larger NDF in the necessary size extrapolation to the $L\rightarrow \infty $ limit in order to reduce and control the error bar. Indeed, we could get the NDF as large as 43. Making use of the obtained large NDF, we carefully examined the stability of our estimates of $T_{{\rm CG}}$ and $T_{{\rm SG}}$.

 Finally, we succeeded in obtaining rather stable and accurate estimates of the CG and SG transition temperatures as $T_{{\rm CG}}=0.142\pm 0.001$ and $T_{{\rm SG}}=0.131^{+0.001}_{-0.006}$. The results provide strong numerical support for the spin-chirality decoupling. The relative magnitude of the finite-size CG correlation length $\xi_{{\rm CG},L}$ and the corresponding SG one $\xi_{{\rm SG},L}$ was also studied. It was found that, on increasing $L$, $\xi_{{\rm CG},L}$  progressively outgrew $\xi_{{\rm SG},L}$ at low temperatures. As already shown in Ref.\cite{HukushimaKawamura2005}, the CG correlation time outgrew the SG correlation time for larger sizes and at lower temperatures. Hence, for larger systems, the CG correlation exceeds the SG correlation both in time and length at lower temperatures, strongly suggesting the occurrence of the spin-chirality decoupling. We also determined the critical exponents associated with the CG and SG transitions. For the CG transition, we got the CG exponents, $\nu_{{\rm CG}}=1.36\pm 0.10$ and $\eta_{{\rm CG}}=0.49\pm 0.10$, consistently with the earlier reports and with the corresponding experimental values on canonical SG. This agreement gives support to the chirality scenario of the experimental SG ordering. The one-step-like feature of the CG ordering reported earlier was also confirmed for larger sizes than before.

\subsection{Relation to other numerical simulation}

 As was already introduced, a simultaneous spin and chiral transition was claimed in several MC works on the same model \cite{Campos,LeeYoung2007,Fernandez,Nakamura}, and we wish to discuss and summarize here how those authors reached a different conclusion from our present conclusion in order to further clarify the situation. Since the system size needs to exceed the crossover-length scale of, say, $15\sim 20$, we take up in the following only large-scale MC simulations with their maximum size $L\geq 32$.

 Campos {\it et al\/} computed both the CG and SG correlation-length ratios up to $L=32$, and observed that the chiral $\xi_{{\rm CG}}/L$ curves cross at temperatures which are only weakly $L$-dependent while the spin $\xi_{{\rm SG}}/L$ curves cross at progressively lower temperatures as $L$ increases \cite {Campos}. Campos {\it et al\/} claimed that the chiral and spin sectors undergo simultaneously a Kosterlitz-Thouless (KT) transition with massive logarithmic corrections. However, the investigated temperature range $T\gtrsim 0.144$ was limited to only above $T_{{\rm CG}}$, which largely restricted the information available in discussing the ordering. The analysis and the interpretation of a simultaneous spin and chiral KT-type transition with massive logarithmic corrections was criticized in Refs.\cite {CampbellKawamura,VietKawamuraPRB}. % Meanwhile, their data themselves including both the spin and the chirality seem fully consistent with our data.

 Lee and Young also computed the same quantities up to $L=32$ down to lower temperatures $T\geq 0.121$ and observed a marginal behavior both for the spin and the chirality \cite{LeeYoung2007}. Comparison of their $\xi_{{\rm CG}} /L$ and  $\xi_{{\rm SG}} /L$ data of $L=32$ with our corresponding data reveals that their $L=32$ data were smaller than ours by about $4\sigma$ for the chirality (in unit of our $\sigma$) and  by about 3$\sigma$ for the spin (their $L=24$ data agree with our data both for the chirality and the spin). We then suspect that the marginal behavior reported in Ref.\cite{LeeYoung2007} was simply due to the too small value of $\xi/L$ for their largest size $L=32$ at low temperatures.

 Fernandez {\it et al\/} computed and analyzed $\xi_{{\rm CG}}/L$ and $\xi_{{\rm SG}}/L$ up to the size $L=48$, and claimed a simultaneous spin and chiral transition at $T_{{\rm CG}}=T_{{\rm SG}}=0.120^{+0.010}_{-0.100}$ \cite{Fernandez}. In fact, when they tried to fit their $\xi_{{\rm CG}}/L$ data to locate $T_{{\rm CG}}$, they obtained an unphysical negative $T_{{\rm CG}}$-value (when we repeat the same fit by using their data, we indeed get $T_{{\rm CG}}=-4\pm 600$). The $T_{{\rm CG}}(=T_{{\rm SG}}$)-value reported in Ref.\cite{Fernandez}, $0.120^{+0.010}_{-0.100}$, was obtained from both $\xi_{{\rm CG}}/L$ and $\xi_{{\rm SG}}/L$ by forcing $T_{{\rm CG}}=T_{{\rm SG}}$, excluding the possibility of the spin-chirality decoupling. So, nothing can actually be concluded concerning the presence or absence of the spin-chirality decoupling from their data themselves. As was examined in detail in \S IIIC, this largely ambiguous situation was originated from their $L=48$ $\xi_{{\rm CG}}/L$ data which was smaller than our data by modest amount, only by $1\sim 1.5\sigma$, the associated crossing temperatures being smaller than ours by $2\sim 3\sigma$. These relatively minor difference, though they themselves are not necessarily contradiction, are combined with the small NDF=1 available in their data fit, and eventually lead to an almost meaningless fitting result, $T_{{\rm CG}}=-4\pm 600$. By contrast, their $L=48$ spin $\xi_{{\rm SG}}/L$ data agree well with ours together with the associated $T_{{\rm cross}}$ values, and their estimate of $T_{{\rm SG}}=0.129^{+0.003}_{-0.016}$ agrees well with our present estimate of $T_{{\rm SG}}=0.131^{+0.001}_{-0.006}$.

 Nakamura recently performed the nonequilibrium MC study on the same model up to the larger size of $L=256$, and claimed a simultaneous spin and chiral transition at $T_{{\rm SG}}=T_{{\rm CG}}=0.140\pm 0.002$ \cite{Nakamura}. Although the nominal size was very large, the probed length was much shorter than $L$. In this nonequilibrium method, the system was quenched from the high temperature, and the subsequent time growth of the SG and CG correlations were analyzed. In this method, the system at lower temperatures of our interest was not fully thermalized at any finite time $t$ and at any length scale. In order to get access to equilibrium properties, one needs to take the long-time limit $t\rightarrow \infty$. As was criticized in detail in Ref.\cite{HukushimaKawamura2005}, however, safely taking the long-time limit $t\rightarrow \infty$ toward equilibrium in hard-relaxing systems like SG is quite a tough issue, and appears to be uncontrolled in this nonequilibrium method. In Ref.\cite{Nakamura}, Nakamura computed the `dynamical correlation length' defined via the time-dependent nonequilibrium CG and SG correlation functions, and observed that, at the estimated simultaneous spin and chiral transition temperature $T=0.140$, the CG one grew only to $\sim 5$ lattice spacings during the nonequilibrium MC simulation, while the equilibrium CG correlation length should eventually diverge there. This length is much shorter than the finite-size correlation lengths $\xi_{{\rm CG},L}$ in full equilibrium, which were estimated in the present paper with the help of the scaling form Eq.(\ref{scaling-xiCG}) as $\xi_{{\rm CG},48}\simeq 20$ for our largest size $L=48$, and as $\xi_{{\rm CG},256}\simeq 130$ for the size of Ref.\cite{Nakamura} $L=256$. The extrapolation from the value of Ref.\cite{Nakamura}, $\xi^{{\rm dynamical}}_{{\rm CG}}\simeq 5$, to the equilibrium value $\xi_{{\rm CG}}\simeq 130$ (and eventually to $\infty$ in the $L\rightarrow \infty$ limit) is highly nontrivial, especially in the situation where the crossover length into the spin-chirality decoupling regime, $15\sim 20$, is to be expected in between.

 Nakamura mentioned that the equilibrium simulation spent most of the CPU time in thermalizing the lattice boundary region which was unnecessary and inefficient, claiming the superiority of the nonequilibrium method \cite{Nakamura}. We disagree with such a view. In nonequilibrium simulations, the dynamical correlation grows from zero after the quench, {\it i.e.\/}, each different region completely uncorrelated initially, to a certain finite length $\ell$ at time $t$, indicating that the size-$\ell$ block is somehow correlated or frozen in its interior. But this does not necessarily mean that the system is fully equilibrated even on the length scale $\lesssim \ell$. Assembly of blocks of size $\ell$ actually forms an interacting network and their states need to be optimized by adjusting to the randomly frustrated interaction between blocks, leading to the block of the size $2\ell$. Such an adjustment or optimization among size-$\ell$ blocks would occur primarily via the interface between these blocks, but it necessarily also gets back to the interior of each size-$\ell$ block modifying the state even within the size-$\ell$ blocks. This procedure continues as $\ell\rightarrow 2\ell\rightarrow 3\ell\rightarrow \cdots$. Of course, blocks and the associated boundaries are somewhat arbitrary inside the lattice for homogenously random systems like the present SG model. Such boundaries between blocks become most eminent at the lattice surface of the length scale $L$ (no free space left), and this might be the reason why the long CPU-time appears to be required to thermalize the lattice-boundary region, as Nakamura noticed. However, essentially similar thermalization process is likely to be occurring at any time and at any length scale $\ell$ everywhere on the lattice, not just at the lattice boundary, but is somehow obscured inside the lattice. To control such a nontrivial thermalization process at all length scales in systems with frustration and randomness constitutes the toughest but most essential part of the SG problem. As such, extreme care needs to be taken in the $t\rightarrow \infty$ limit in the nonequilibrium method in hard-relaxing systems like SG. In our opinion, the orthodox way to go is, after all, to perform an equilibrium simulation on finite-size systems and carefully control the finite-size effect, always fully equilibrating the system at each size.

\subsection{Other related systems}

 Now, we wish to discuss several related systems with some possible relevance to the present issue, especially the spin-chirality decoupling. Our first example is the 3D {\it XY\/} SG with two-component vector spins, which has also been studied for years. It has relevance not only to easy-plane-type random magnets but also to ceramic superconductors with anisotropic pairing symmetry such as cuprates \cite{Kawamurabook,Kawamurareview,KawamuraLi-ceramic}. This system has many common features with the 3D Heisenberg SG. Especially, it possesses a nontrivial $Z_2$ chiral degrees of freedom where the chirality is defined by the vector product of the two neighboring {\it XY\/} spins as $[S_i\times S_j]_z$. In contrast to the chirality for the Heisenberg spin which is cubic in the spin variables and time-reversal odd, the chirality for the {\it XY\/} spin is quadratic in the spin variables and time-reversal even.

 The research on the 3D {\it XY\/} SG followed more or less similar path to that of the Heisenberg SG. Initial studies including both MC simulation \cite{JainYoung} and  numerical domain-wall RG calculation \cite{Morris} indicated that the 3D {\it XY\/} SG did not exhibit any finite-temperature transition. In 1991, Kawamura and Tanemura suggested that the system might exhibit a finite-temperature transition in its {\it chiral\/} sector even though the conventional SG transition occurred only at $T=0$, {\it i.e.\/}, the spin-chirality decoupling \cite{KawamuraTanemura1991}. Since then, many works have been done on this  model with particular interest in the presence or absence of the spin-chirality decoupling. Mentioning some of large-scale MC simulation on the model, Kawamura and Li simulated the model with the $\pm J$ coupling of the lattice size $L\leq 16$, and observed the spin-chirality decoupling, {\it i.e.\/}, $T_{{\rm CG}}>T_{{\rm SG}}$, together with the one-step-like RSB for the chirality \cite{KawamuraLi}. By contrast, Pixley and Young simulated the model with the Gaussian coupling up to the size $L=24$, and claimed a simultaneous spin and chiral transition \cite{PixleyYoung}. Obuchi and Kawamura simulated the same model extending the lattice size up to $L=40$, and reported the spin-chirality decoupling with the estimates, $T_{{\rm CG}}=0.313^{+0.013}_{-0.018}$ and $T_{{\rm SG}}=0.275^{+0.013}_{-0.052}$ \cite{ObuchiKawamura}. These authors also estimated both the CG and SG critical exponents at each transition, $\nu_{{\rm CG}}=1.36^{+0.15}_{-0.37}$ and $\eta_{{\rm CG}}=0.26^{+0.29}_{-0.26}$ for the CG transition, and  $\nu_{{\rm SG}}=1.22^{+0.26}_{-0.06}$ and $\eta_{{\rm SG}}=-0.54^{+0.24}_{-0.52}$ for the SG transition. These exponent values are not far from the corresponding exponents of the 3D Heisenberg SG determined in the present paper, and the relation between the two poses an interesting question.

 Our second example is the Heisenberg SG in different spatial dimensions, {\it e.g.\/}, in 2D and in 4D. In 2D, both the spin and the chirality order only at $T=0$. Yet, the CG correlation-length exponent $\nu_{{\rm CG}}$ appears to be larger than the SG one $\nu_{{\rm SG}}$ by factor of about two \cite{KawamuraYonehara}. Although the transition temperatures are common at $T=0$ for the spin and the chirality, this actually means the spin-chirality decoupling in the sense that this $T=0$ transition possesses two distinct length scales, violating the standard one-length scaling. In 3D, our present work has shown that the spin and the chirality are decoupled, with mutually distinct nonzero transition temperatures $T_{{\rm CG}}>T_{{\rm SG}}>0$. In 4D, by contrast, Kawamura and Nishimura suggested that the spin and the chirality appeared to be {\it coupled\/} in the sense that both ordered at a common finite temperature with a common correlation-length exponent $\nu_{{\rm CG}}\simeq \nu_{{\rm SG}}=1.0\pm 0.1$ \cite{KawamuraNishikawa}.

 Thus, the spin and chiral ordering of the Heisenberg SG depends heavily on its spatial dimensionality. For high enough dimensions $d\geq 4$, the spin-chirality decoupling seems not to occur, suggesting that fluctuation effects are crucial to realize it. On the other hand, if fluctuations are too strong, they simply wash out any finite-temperature transition. In this sense, 3D happens to be the most interesting case where the strength of fluctuations are balanced to realize the spin-chirality decoupling at finite temperatures.

 In this connection, it might be useful to point out that similar spin-chirality decoupling phenomenon has been discussed in {\it regularly frustrated XY\/} model in 2D. For this model, after some controversy, it now exists a consensus that the spin-chirality decoupling indeed occurs, {\it i.e.\/}, the chirality orders at a temperature slightly higher than that of the spin, though only with a small difference of 1\% order \cite{Okumura,ObuchiKawamura-triangular}. Since there is an empirical observation that the random system often behaves similarly to the corresponding regular system with less spatial dimension (the so-called dimensional reduction), the occurrence of the spin-chirality decoupling in 3D might not be so surprising in view of the now established spin-chirality decoupling in 2D regularly frustrated systems.

Our last example has an apparent connection with the above-mentioned systematic variation of the SG and CG ordering behavior with respect the spatial dimensionality $d$. An interesting observation has been made on certain 1D SG model with long-range power-law interaction, $J_{ij}\propto 1/r_{ij}^\sigma$, that varying the power of the long-range interaction $\sigma$ might correspond to varying the space dimensionality $d$ of the short-range model. If this is the case, since the 1D model can be simulated to very large lattice size $L$ even under the long-range interaction, it might shed further light to the issue of the SG and CG order in $d$-dimensional Heisenberg SG.

 Mentioning some of large-scale MC simulations on the 1D Heisenberg SG model with the long-range power-law interaction, Viet and Kawamura simulated random exchange model with the Gaussian coupling up to the size $L=4096$, to find that this 1D model exhibited the spin-chirality decoupling of $T_{{\rm CG}}>T_{{\rm SG}}\geq 0$ in the range $0.8\lesssim \sigma\lesssim 1.1$ \cite{VietKawamura1D}. Sharma and Young dealt with a different type of randomness where the interaction between distant spins were diluted according to the power law. By simulating the model up to the size $L=16384$, the authors observed the spin-chirality decoupling $T_{{\rm CG}}>T_{{\rm SG}}>0$ at $\sigma=0.85$ \cite{SharmaYoung}.

 In this way, the problem of the spin and chiral ordering in the 3D Heisenberg SG has rich connections to many other problems and systems.

\subsection{Relation to experiments}

Finally, we wish to turn to our original problem, {\it i.e.\/}, the chirality scenario of the ordering of real Heisenberg-like SG. The crucial ingredient of the scenario is the spin-chirality decoupling in the hypothetical fully isotropic Heisenberg SG. Weak random magnetic anisotropy, which inevitably exists in real SG magnets, ``recouples'' the spin into the chirality, and the CG transition and the CG order manifest themselves as the SG transition and the SG order via the weak random magnetic anisotropy.

 The scenario already got some experimental supports. The first support comes from the critical exponents of Heisenberg-like SG. The chirality scenario predicts that the SG exponents of weakly anisotropic Heisenberg-like SG including canonical SG are nothing but the CG exponents of the hypothetical fully isotropic Heisenberg SG \cite{Kawamurabook,Kawamurareview}. Here, for completeness, we elaborate how such an equivalence of the experimentally-observable SG exponents of real Heisenberg-like SG magnets to the CG exponents of the fully isotropic Heisenberg SG is to be expected in the framework of the chirality scenario.

 In the fully isotropic case, the system is invariant under both the proper spin rotation $SO(3)$ and the time-reversal (the spin-inversion) $Z_2$, ${\bm S}_i \rightarrow - {\bm S}_i$. While the Heisenberg spin ${\bm S}_i$ is accordingly transformed under the both operations, the chirality $\chi$ remains invariant under $SO(3)$ but changes its sign under $Z_2$. Let us introduce the pseudospin $\tilde {\bm S}_i$ by the relation ${\bm S}_i=\chi_i \tilde {\bm S}_i$. Though $\tilde {\bm S}_i$ transforms as a vector under $SO(3)$, it remains invariant under $Z_2$ in contrast to the original spin ${\bm S}_i$. The spin-chirality decoupling then means that $SO(3)$ and $Z_2$ are decoupled on long length scales, leading to separate CG and SG transitions, each associated with the spontaneous breaking of the $Z_2$ and $SO(3)$ symmetries. Under the spin-chirality decoupling, the SG correlation function would behave in the long-distance limit as
\begin{eqnarray}
g_{{\bm S}}({\bm r}_{ij}) &\equiv& [\langle {\bm S}_i \cdot {\bm S}_j \rangle^2] \nonumber \\
 &=& [\langle \chi_i \tilde{\bm S}_i \cdot \chi_j \tilde{\bm S}_j \rangle^2] \nonumber \\
 &\simeq& [\langle \chi_i \chi_j\rangle^2]\times [\langle \tilde{\bm S}_i \cdot \tilde{\bm S}_j \rangle^2] \nonumber \\
 &=& g_\chi ({\bm r}_{ij}) g_{\tilde{\bm S}}({\bm r}_{ij}).
\label{correlation}
\end{eqnarray}
In the CG ordered state of the isotropic Heisenberg SG, the chiral $g_\chi$ takes a nonzero value even at $|{\bm r}_{ij}|\rightarrow \infty$, while the spin $g_{\tilde{\bm S}}$ decays to zero since $g_{\tilde{\bm S}}$ decays to zero due to the unbroken $SO(3)$, and the standard SG order does not arise. In the presence of the weak random magnetic anisotropy which inevitably exists in real Heisenberg-like SG, the Hamiltonian symmetry reduces from $Z_2\times SO(3)$ to only $Z_2$, no longer $SO(3)$ symmetry left. The chirality scenario then claims that the only thermodynamic transition occurring in such weakly anisotropic system is the one associated with the $Z_2$-symmetry breaking, {\it i.e.\/}, the CG one, at which the SG order is simultaneously induced parasitic to the CG order via the random magnetic anisotropy reflecting the absence of the $SO(3)$ symmetry (``{\it spin-chirality recoupling\/}'' \cite{Kawamurabook,Kawamurareview}). The character of the $Z_2$-symmetry breaking would essentially be the same as that of the CG transition of the fully isotropic system, since the broken symmetry, chiral $Z_2$, is the same, with $SO(3)$ being decoupled. Due to the absence of the $SO(3)$ symmetry at the Hamiltonian level, $g_{\tilde{\bm S}}$ in the long-distance limit takes a nonzero constant. Hence, at the CG (and simultaneously SG) transition of the weakly anisotropic SG, $g_{\tilde{\bm S}}$ takes a nonzero value $A$ at $|{\bm r}|\rightarrow \infty$ even at and above the transitin temperature, while $g_\chi$ exhibits the critical behavior described by the CG exponents of the isotropic system, {\it e.g.\/}, $\nu_{{\rm CG}}$ and $\eta_{{\rm CG}}$ as determined in the present paper. Then, one expects from Eq.(\ref{correlation}),
\begin{eqnarray}
g_{{\bm S}}({\bm r}_{ij}) \simeq A g_\chi ({\bm r}_{ij}), \ \ \ \ {\rm for}\ |{\bm r}_{ij}|\rightarrow \infty ,
\label{correlation2}
\end{eqnarray}
indicating that the experimentally-observable {\it spin\/} critical properties of the realistic Heisenberg-like SG magnets like canonical SG should be those of the CG critical properties of the fully isotropic system.

 The critical properties of canonical SG are well-studied, various measurements providing mutually consistent estimates, $\alpha\simeq -2.0$, $\beta\simeq 1.0$, $\gamma\simeq 2.0-2.2$, $\delta\simeq 3.0-3.3$, $\nu\simeq 1.3-1.4$ and $\eta\simeq 0.4$ \cite{Kawamurabook}. These values are indeed quite close to our present estimates of the CG exponents $\alpha\simeq -2.1$, $\beta\simeq 1.0$, $\gamma\simeq 2.1$, $\delta\simeq 3.0$, $\nu\simeq 1.36$ and $\eta\simeq 0.49$.

 If ones compares the experimental exponents with the SG exponents of the decoupled SG transition estimated in the present paper, $\alpha\simeq -1.6$, $\beta\simeq 0.45$, $\gamma\simeq 2.7$, $\delta\simeq 7.0$, $\nu\simeq 1.2$ and $\eta\simeq -0.25$, some exponents are not far, but some other exponents, {\it e.g.\/}, $\beta$, $\delta$ and $\eta$, differ considerably. Hence, the decoupled SG transition of the Heisenberg SG, even though it occurs at nonzero temperature, cannot explain the experimental exponents of Heisenberg-like SG properly. By contrast, the chirality scenario is capable of explaining the experimental exponents quantitatively as the CG ones.  We also note that the exponents of the 3D Ising SG, which were believed in the earlier view to govern the asymptotic criticality of the weakly anisotropic Heisenberg-like SG including cannical SG \cite{BrayMooreYoung}, are also far from the experimental exponent values, {\it e.g.\/}, $\gamma$ of the 3D Ising SG is $\gamma=6.0\sim 6.5$ in contrast to the experimental value of canonical SG, $\gamma\simeq 2$ \cite{Kawamurabook}. By contrast, for the Ising-like SG magnet Fe$_x$Mn$_{1-x}$TiO$_3$, the experimentally observed exponent,  $\gamma\simeq 4$, is much larger than that of canonical SG and is indeed close to the theoretically determined value of the 3D Ising SG model \cite{Kawamurabook,Campbellreview}.

 The second and most direct experimental evidence of the chirality scenario comes from the Hall measurements. Experimentally, the measured Hall coefficient of canonical SG, the Hall resistivity divided by the magnetization, exhibits a cusp-like anomaly at the SG transition \cite{Campbell2004,Taniguchi2004,Campbell2006,Taniguchi2007,Yamanaka2007}. The effect is neither the normal Hall effect nor the usual anomalous Hall effect, and can be ascribed to the {\it topological Hall effect\/} originated from the chiral order. In fact, the Hall coefficient was shown to correspond to the `chiral susceptibility' of canonical SG \cite{TataraKawamura,KawamuraHall}, and the detection of the strong singularity is a direct experimental demonstration of the CG order. Estimating the CG critical exponents from the Hall signal is a challenging task. The only report so far made concerns with $\delta$. Taniguchi reported $\delta=2.5\pm 0.8$ \cite{Taniguchi2007}, which is consistent with the value expected from the chirality scenario $\delta_{{\rm CG}}\simeq 3$. Experimental determination of other critical exponents from the Hall measurements would be highly desirable for the future task.
% Hall measurements on the canonical SG showing the reentrant transitions, {\it i.e.\/}, successive transitions first from the paramagnetic to the ferromagnetic states at $T_{c1}$, and then from the ferromagnetic state to the SG state at $T_{c2}(<T_{c1})$. The measured Hall resistivity exhibits sharp anomalies both at $T_{c1}$ and $T_{c2}$, but the Hall coefficient exhibits an anomaly only at $T_{c2}$. This clearly demonstrates that the anomaly at $T_{c2}$ at the SG transition arises not from the standard anomalous Hall effect proportional to the magnetization, but the topological one.

 The third experimental evidence of the chirality scenario concerns with the in-field SG ordering and the magnetic phase diagram of Heisenberg-like SG \cite{Kawamurareview,Kawamurabook,Orbach}. Often, in-field transition lines appear in the phase diagram in the magnetic field ($H$) versus temperature ($T$) plane, which have widely been interpreted in terms of the mean-field (MF) theory and are usually called the `AT line' and the `GT line', the former sometimes further divided into the `low-field AT line' and the `high-field AT line' \cite{Orbach}. The behavior of the GT and AT phase boundaries in the $H$-$T$ phase diagram, $H\approx |T-T_g|^x$, is characterized by the exponents $x=1/2$ and 3/2, respectively. While the experimentally observed exponents describing the behavior of the in-field phase boundary agree well with the corresponding MF exponents, the coefficient of the low-field AT line has been known to largely deviate from the MF value, say, by factor of 30, whereas no such deviation is observed for the GT line and the high-field AT line \cite{Kawamurareview,Kawamurabook,Orbach}. This failure of the MF theory remains to be explained. In addition, concerning the exponent values themselves, the MF theory usually gives poor numerics, and the reason why it gives rather accurate exponent values only in the case of the in-field phase boundaries sounds a bit odd (for example, the zero-field critical exponents such as $\gamma$ largely deviate from the MF value as usual, {\it i.e.\/}, $\gamma\simeq 2$ in canonical SG versus $\gamma=1$ in MF). The chirality scenario presents a completely different explanation of the apparently MF-like behavior of the in-field phase boundaries \cite{Kawamurareview,Kawamurabook}. The scenario is also capable of explaining the deviation of the coefficient of the low-field AT-line \cite{Kawamurareview}. Further details about the magnetic phase diagram of the Heisenberg-like SG and the chirality scenario will be published elsewhere \cite{Kawamura-preparation}. 

 Finally, as the fourth experimental support of the chirality scenario, we wish to touch upon the RSB structure of the SG ordered state. In SG magnets {\it below\/} $T_g$, while full thermalization is usually not possible, various intriguing off-equilibrium phenomena such as aging arise there \cite{Kawamurabook,Vincentreview}. In such off-equilibrium situation, it has been realized that the breaking pattern  of the fluctuation-dissipation relation held in equilibrium gives information about the RSB pattern of the SG order. Indeed, the seminal experiment by H\'erisson and Ocio on Heisenberg-like insulating SG thiospinel CdCr$_{1.7}$In$_{0.3}$S$_4$ gave the information on its RSB pattern \cite{HerissonOcio2002,HerissonOcio2004}. Due to the difficulty in taking the long waiting-time limit, the result was not necessarily conclusive, but it favored a one-ste-like RSB \cite{HerissonOcio2004} as expected in the chirality scenario. On theoretical side, the off-euilibrium MC simulation on the weakly anisotropic 3D Heisenberg-like SG model of Ref.\cite{Kawamura2003} gave some evidence of the one-step-like feature of the `fluctuation-dissipation ratio', consistently with the chirality scenario. Furthermore, the experimentally observed `effective temperature' of the SG state, $\sim 1.9T_g$, \cite{HerissonOcio2002,HerissonOcio2004} is rather close to the corresponding theoretical value of $\sim 2T_g$ \cite{Kawamura2003}.

 Thus, the chirality scenario provides a quite promising framework in systematically understanding the ordering of Heisenberg-like SG. The spin-chirality decoupling in the fully isotropic Heisenberg SG is the basis of the scenario, and our present result strengthens the validity of the hypothesis.

 SG is a hard-relaxing system, and one might wonder if its full thermalization might be limited to rather short length scale even above $T_g$ in real SG. Since the spin-chirality decoupling is a phenomenon {\it in equilibrium\/} arising over the crossover-length scale of $15\sim 20$ lattice spacings, the target SG system needs to be thermalized at least over this length scale above $T_g$. In this connection, L\'evy and Ogielski reported for canonical SG AgMn that the equilibrium SG critical properties were measurable above $T_g$ on the length scale of a few hundreds lattice spacings \cite{Levy}. This is far more than the crossover length scale, indicating that the spin-chirality decoupling could play a role and could be `seen' in real SG ordering. 

  From the computational viewpoint, it seems not so easy within the present methodology to go much beyond $L=48$ in {\it equilibrium\/} simulations. We most probably need some new methodology to go much beyond this size. On the other hand, to fill the gap with experiments, {\it i.e.\/}, increasing the present $L\simeq 48$ toward a few hundreds {\it keeping equilibrium\/} has a real physical signicance, not just an academic issue. Thus, further numerical efforts in this direction would deserve a serious challenge.

\begin{acknowledgments}
The authors are thankful to K. Hukushima, Y. Tabata, E. Vincent, H. Yoshino and T. Taniguchi for helpful discussion. This study was supported by JSPS KAKENHI Grant Number JP17H06137. We thank ISSP, Tokyo University, YITP, Kyoto University, and CMC, Osaka University, for providing us with the CPU time.
\end{acknowledgments}

\end{document}